\documentclass[sigconf]{acmart}
\usepackage{booktabs} 
\usepackage{tabularx}
\usepackage{graphicx}
\usepackage{url}
\usepackage{mdwlist}
\usepackage{subcaption}
\usepackage{hyperref}
\hypersetup{colorlinks=false,linkcolor=blue,urlcolor=blue,citecolor=red}
\usepackage{amsfonts}
\usepackage[labelfont=bf,textfont=bf]{caption}
\usepackage{multirow}
\usepackage{array}
\usepackage[linesnumbered,ruled]{algorithm2e}
\SetKwComment{Comment}{$\triangleright$\ }{}
\usepackage{xcolor}
\setcopyright{rightsretained}
\usepackage{amssymb}
\usepackage[flushleft]{threeparttable}
\usepackage{enumitem}
\usepackage{balance}

\newcommand{\ie}{\emph{i.e., }}
\newcommand{\eg}{\emph{e.g., }}
\newcommand{\etal}{\emph{et al.}}

\newcommand{\aka}{\emph{a.k.a. }}

\begin{document}
	
\copyrightyear{2018} 
\acmYear{2018} 
\setcopyright{acmcopyright}
\acmConference[SIGIR'18]{41st International ACM SIGIR Conference on Research and Development in Information Retrieval}{July 8--12, 2018}{Ann Arbor, MI, USA}
\acmBooktitle{SIGIR '18: 41st International ACM SIGIR Conference on Research and Development in Information Retrieval, July 8-12, 2018, Ann Arbor, MI, USA}
\acmPrice{15.00}
\acmDOI{10.1145/3209978.3209981}
\acmISBN{978-1-4503-5657-2/18/07}
\fancyhead{}

\title{Adversarial Personalized Ranking for Recommendation}
\titlenote{
	This work is done when during the internship of Zhankui He and Xiaoyu Du at National University of Singapore. 
	NExT research is supported by the National Research Foundation, Prime Minister's Office, Singapore under its IRC@SG Funding Initiative.}

\author{Xiangnan He}
\affiliation{%
	\institution{National University of Singapore}
}
\email{xiangnanhe@gmail.com}

\author{Zhankui He}
\affiliation{%
	\institution{Fudan University}
}
\email{zkhe15@fudan.edu.cn}

\author{Xiaoyu Du}
\affiliation{%
	\institution{Chengdu University of Information Technology}
}
\email{duxy.me@gmail.com}

\author{Tat-Seng Chua}
\affiliation{%
	\institution{National University of Singapore}
}
\email{dcscts@nus.edu.sg}

\ccsdesc[500]{Information systems~Recommender systems}
\ccsdesc[500]{Information systems~Information retrieval}
\ccsdesc[500]{Information systems~Retrieval models and ranking}


\begin{abstract}
Item recommendation is a personalized ranking task. To this end, many recommender systems optimize models with pairwise ranking objectives, such as the Bayesian Personalized Ranking (BPR). 
Using matrix Factorization (MF) --- the most widely used model in recommendation --- as a demonstration, we show that optimizing it with BPR leads to a recommender model that is not robust. In particular, we find that the resultant model is highly vulnerable to adversarial perturbations on its model parameters, which implies the possibly large error in generalization. 

To enhance the robustness of a recommender model and thus improve its generalization performance, we propose a new optimization framework, namely \textit{Adversarial Personalized Ranking} (APR). In short, our APR enhances the pairwise ranking method BPR by performing adversarial training. It can be interpreted as playing a minimax game, where the minimization of the BPR objective function meanwhile defends an adversary, which adds adversarial perturbations on model parameters to maximize the BPR objective function. 
To illustrate how it works, we implement APR on MF by adding adversarial perturbations on the embedding vectors of users and items. 
Extensive experiments on three public real-world datasets demonstrate the effectiveness of APR --- by optimizing MF with APR, it outperforms BPR with a relative improvement of $11.2\%$ on average and achieves state-of-the-art performance for item recommendation. Our implementation is available at: \url{https://github.com/hexiangnan/adversarial_personalized_ranking}. 



\end{abstract}

\keywords{Personalized Ranking, Pairwise Learning, Adversarial Training, Matrix Factorization, Item Recommendation}
\maketitle

\section{Introduction}
\label{sec:introduction}

Recent advances on adversarial machine learning~\cite{AML_2014} show that many state-of-the-art classifiers are actually very fragile and vulnerable to \textit{adversarial examples}, which are formed by applying small but intentional perturbations to input examples from the dataset. A typical example can be found in Figure 1 of \cite{AML_2015}, which demonstrates that by adding small adversarial perturbations to an image of panda, a well-trained classier misclassified the image as a gibbon with a high confidence, whereas the effect of perturbations can hardly be perceived by human. This points to an inherent limitation of training a model on static labeled data only. To address the limitation and improve model generalization, researchers then developed adversarial training methods that train a model to correctly classify the dynamically generated adversarial examples~\cite{AML_2015,moosavi2016universal}. 

While the inspiring progress of adversarial machine learning mainly concentrated on the computer vision domain where the adversarial examples can be intuitively understood, to date, there is no study about such adversarial phenomenon in the field of information retrieval (IR). Although the core task in IR is ranking, we point out that many learning to rank (L2R) methods are essentially trained by optimizing a classification function, such as the pairwise L2R method Bayesian Personalized Ranking (BPR) in recommendation~\cite{BPR}, among others~\cite{book_l2r}. This means that it is very likely that the underlying IR models also lack robustness and are vulnerable to certain kinds of ``adversarial examples''. 
In this work, we aim to fill the research gap by exploring adversarial learning methods on item recommendation, an active and fundamental research topic in IR that concerns personalized ranking.

Nevertheless, directly grafting the way of generating adversarial examples from the image domain is infeasible, since the inputs of recommender models are mostly discrete features (\ie user ID, item ID, and other categorical variables). Clearly, it is meaningless to apply noises to discrete features, which may change their semantics. To address this issue, we consider exploring the robustness of a recommender model at a deeper level --- at the level of its intrinsic model parameters rather than the extrinsic inputs. Using the matrix factorization (MF) model~\cite{fastMF,SVD++} trained with BPR as a demonstration (we term this instantiation as MF-BPR), we investigate its robustness to perturbations on embedding parameters. Note that MF-BPR is a highly competitive approach for item recommendation and has been used in many papers as the state-of-the-art baseline up until recently~\cite{NCF}. We found that MF-BPR is not robust and is vulnerable to adversarial perturbations on the parameters. This sheds light on the weakness of training with BPR, and motivates us to develop adversarial learning methods that can result in better and more robust recommender models. 

As the main contribution of this work, we propose a novel \textit{Adversarial Personalized Ranking} (APR) method to learn recommender models. With BPR as the building block, we introduce an additional objective function in APR to quantify the loss of a model under perturbations on its parameters. The formulation of APR can be seen as playing a minimax game, where the perturbations are optimized towards maximizing the BPR loss, and the model is trained to minimize both the BPR loss and the additional loss with adversarial perturbations. With a differentiable recommender model, the whole framework of APR can be optimized with the standard stochastic gradient descent. To demonstrate how it works, we derive the APR solver for MF and term the method as \textit{Adversarial Matrix Factorization} (AMF). We conduct extensive experiments on three public datasets constructed from Yelp, Pinterest and Gowalla that represent various item recommendation scenarios. Both quantitative and qualitative analysis justify the effectiveness and rationality of adversarial training for personalized ranking. Specifically, our AMF outperforms MF-BPR with a significant improvement of $11\%$ on average in NDCG and hit ratio. It also outperforms the recently proposed neural recommender models~\cite{NCF,CDAE} and IRGAN~\cite{IRGAN}, and achieves state-of-the-art performance for item recommendation. 




\section{Preliminaries}
\label{sec:preliminary}
First the matrix factorization model for recommendation is described. Next the pairwise learning method Bayesian Personalized Ranking is shortly recapitulated. The novel contribution of this section is to empirically demonstrate that the MF model optimized by BPR (\aka MF-BPR) is not robust and is vulnerable to adversarial perturbations on its parameters. 

\subsection{Matrix Factorization}
MF has been recognized as the basic yet most effective model in recommendation since several years~\cite{iCD,SVD++,DCF}.
Being a germ of representation learning, MF represents each user and item as an embedding vector.
The core idea of MF is to estimate a user's preference on an item as the inner product between their embedding vectors. 
Formally, let $u$ denote a user and $i$ denote an item, then the predictive model of MF is formulated as: 
$
	\hat{y}_{ui}(\Theta) = \textbf{p}_u^T \textbf{q}_i,
$
where $\textbf{p}_u\in\mathbb{R}^K$ and $\textbf{q}_i\in\mathbb{R}^K$ denote the embedding vector for user $u$ and item $i$, respectively, and $K$ is the size of embedding vector also called as \textit{embedding size}. $\Theta$ denotes the model parameters of MF, which is consisted of all user embedding and item embedding vectors, \ie $\Theta=\{\{\textbf{p}_u\}_{u\in \mathcal{U}}, \{\textbf{q}_i\}_{i \in \mathcal{I}} \}$, where $\mathcal{U}$ and $\mathcal{I}$ denote the set of all users and items, respectively. We use $\textbf{P}$ and $\textbf{Q}$ to denote the embedding matrix $\textbf{P}=\{\textbf{p}_u\}_{u\in \mathcal{U}}, \textbf{Q}=\{\textbf{q}_i\}_{i \in \mathcal{I}}$ for short. 




\subsection{Bayesian Personalized Ranking}
BPR is a pairwise L2R method and has been widely used to optimize recommender models towards personalized ranking~\cite{BPR}. Targeting at learning from implicit feedback, it assumes that observed interactions should be ranked higher than the unobserved ones.
To this end, BPR maximizes the margin between an observed interaction and its unobserved counterparts. This is fundamentally different from pointwise methods~\cite{iCD,NCF} that optimize each model prediction towards a predefined groundtruth. Formally, the objective function (to be minimized) of BPR is
\begin{equation}
	L_{BPR}(\mathcal{D}|\Theta) = \sum_{(u,i,j)\in \mathcal{D}} -\ln \sigma(\hat{y}_{ui}(\Theta) - \hat{y}_{uj}(\Theta)) + \lambda_{\Theta} ||\Theta||^2,
\end{equation}
where $\sigma(\cdot)$ is the sigmoid function, $\lambda_{\Theta}$ are model specific regularization parameters to prevent overfitting, and $\mathcal{D}$ denotes the set of pairwise training instances
$
	\mathcal{D} := \{(u,i,j)| i\in\mathcal{I}_u^+ \wedge j\in\mathcal{I} \setminus \mathcal{I}_u^+ \},
$
where $\mathcal{I}_u^+$ denotes the set of items that $u$ has interacted with before, and $\mathcal{I}$ denotes the whole item set. 
Since the number of training instances in BPR is very huge,
the optimization of BPR is usually done by performing stochastic gradient descent (SGD). After obtaining parameters, we can get the personalized ranked list for a user $u$ based on the value of $\hat{y}_{ui}(\Theta)$ over all items.

\begin{figure*}[t]
	\centering
	\begin{subfigure}[b]{0.24\textwidth}
		\centering
		\includegraphics[width=\textwidth]{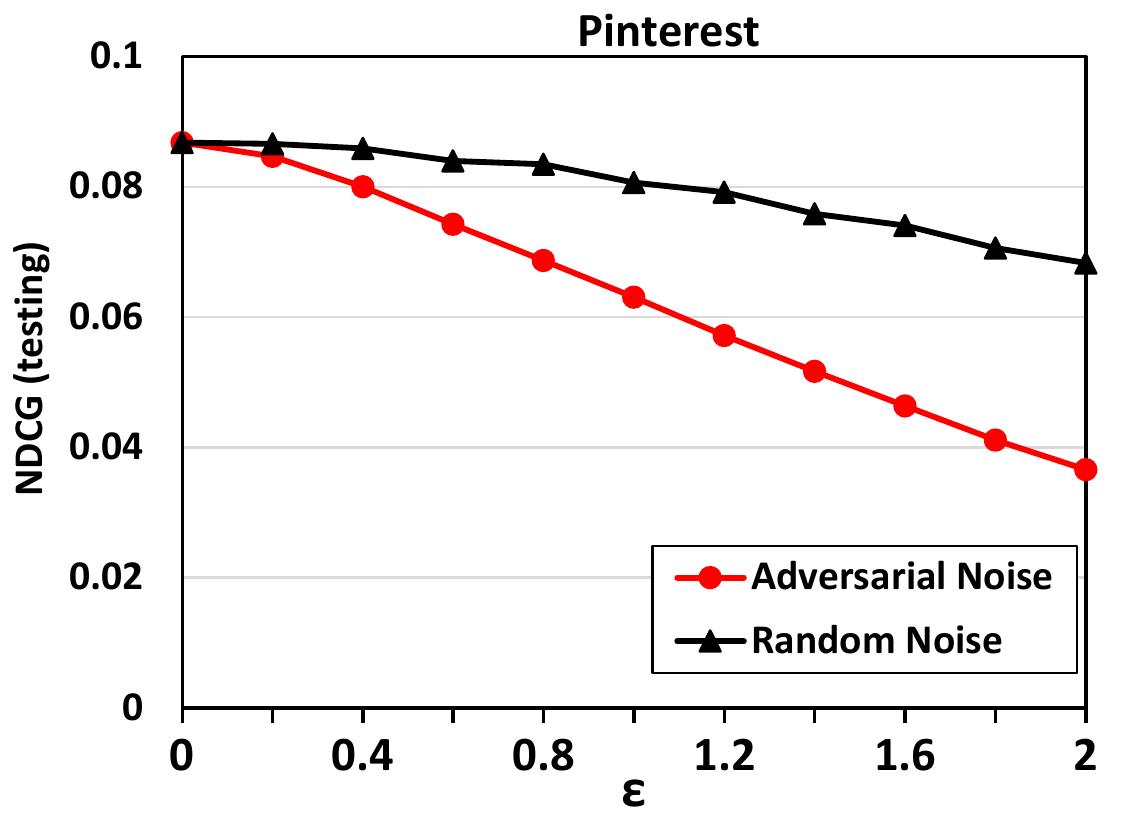}
		\vspace{-18pt}
		\caption{Testing NDCG vs. $\epsilon$}
		\label{fig:pinterest-noise-ndcg}
	\end{subfigure} 
	\begin{subfigure}[b]{0.24\textwidth}
		\centering
		\includegraphics[width=\textwidth]{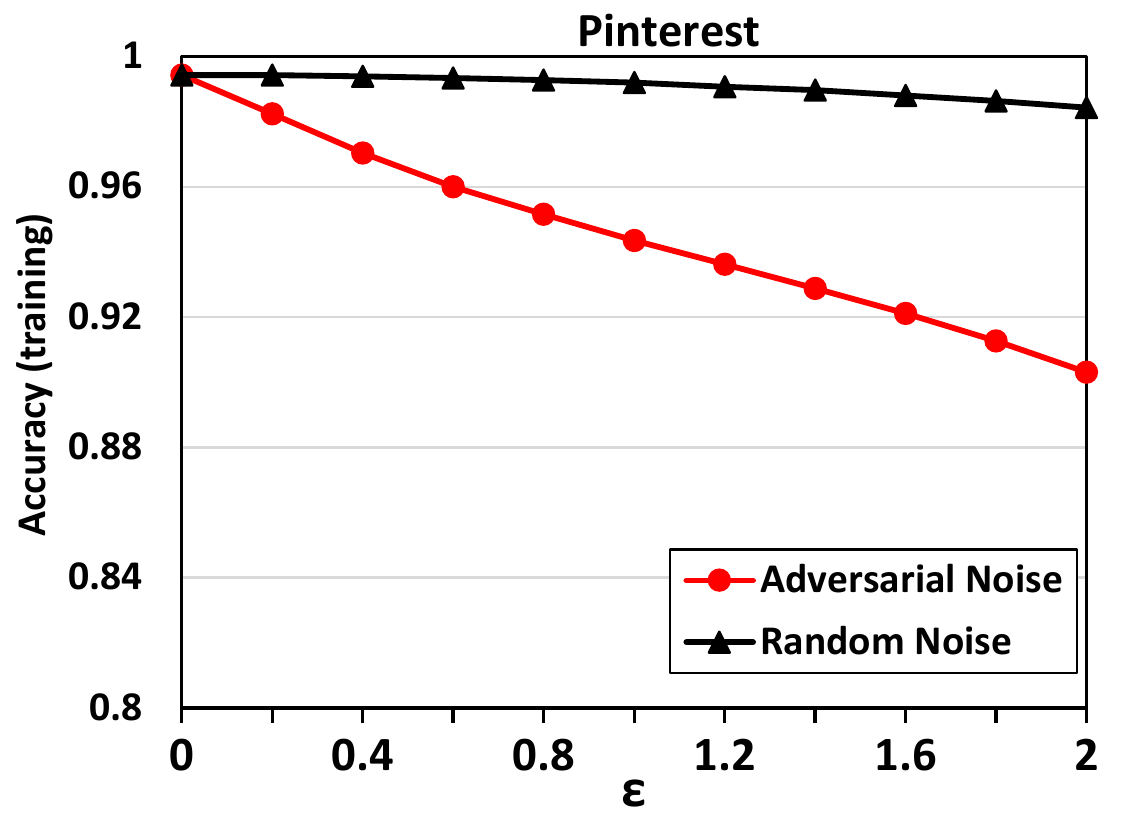}
		\vspace{-18pt}
		\caption{Training Accuracy vs. $\epsilon$}
		\label{fig:pinterest-noise-acc}
	\end{subfigure} 
	\begin{subfigure}[b]{0.24\textwidth}
		\centering
		\includegraphics[width=\textwidth]{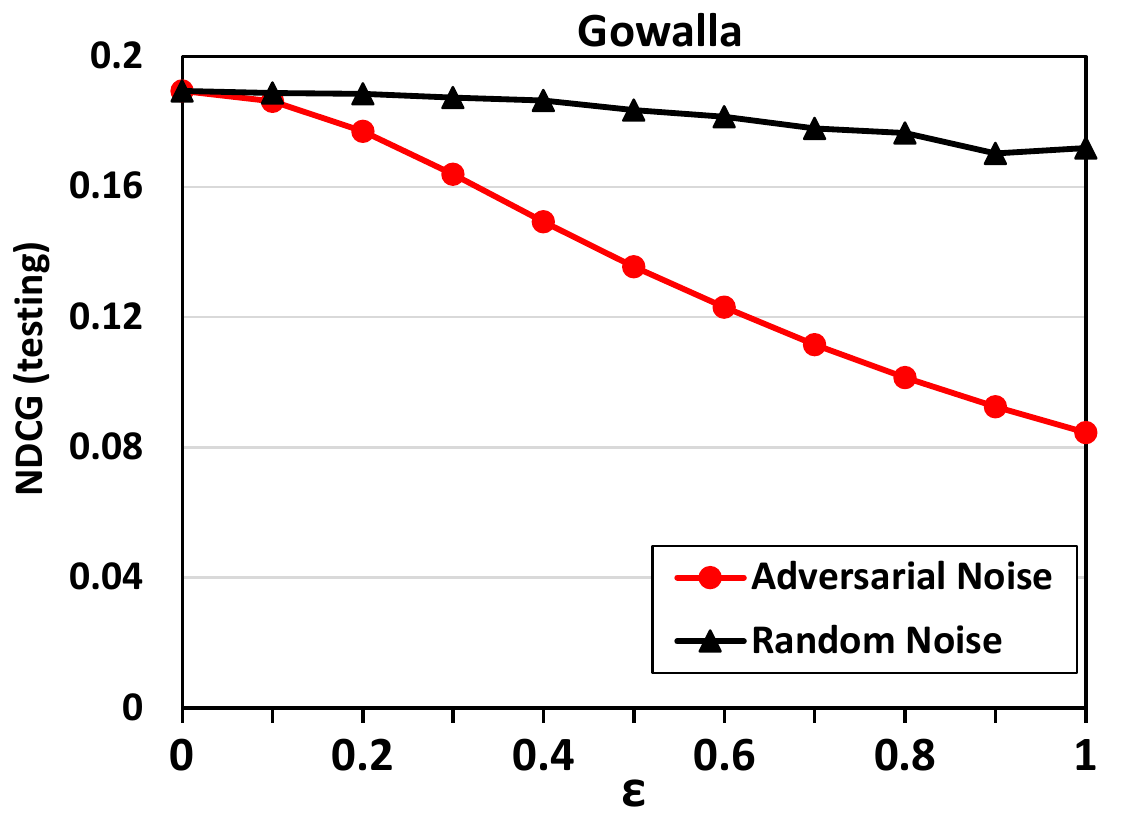}
		\vspace{-18pt}
		\caption{Testing NDCG vs. $\epsilon$}
		\label{fig:gowalla-noise-ndcg}
	\end{subfigure} 
	\begin{subfigure}[b]{0.24\textwidth}
		\centering
		\includegraphics[width=\textwidth]{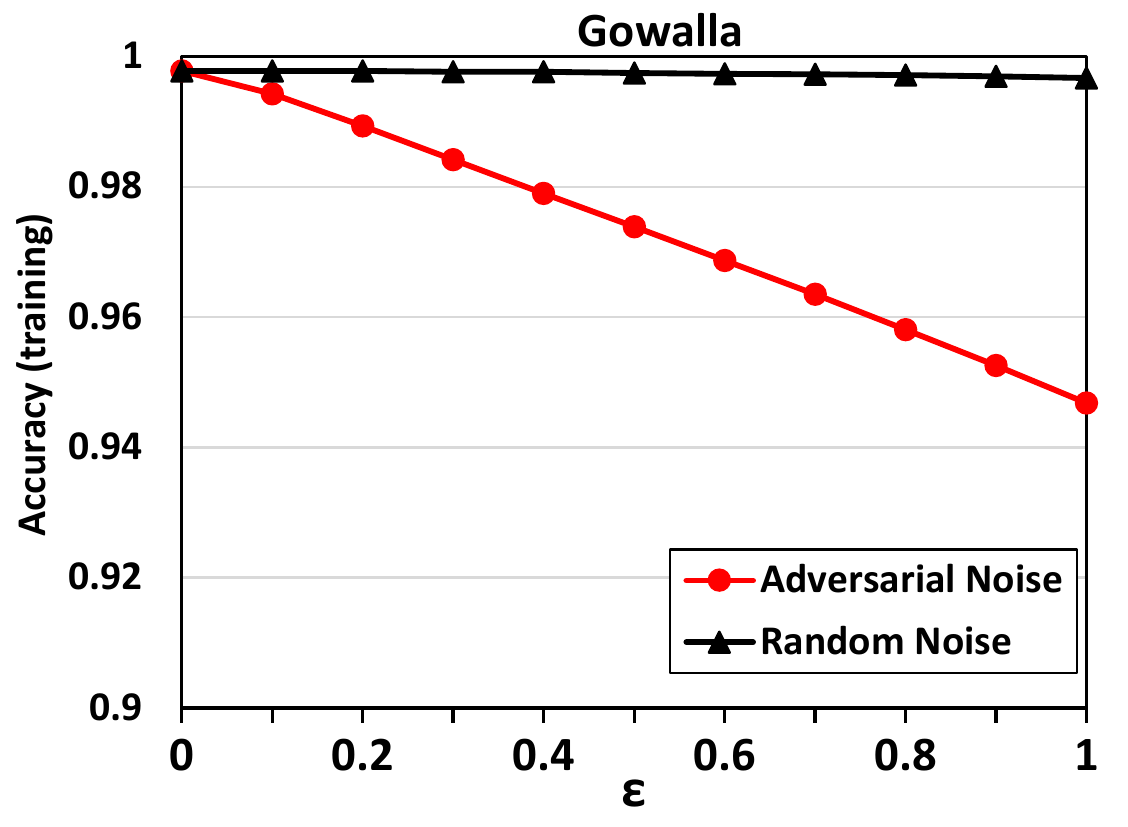}
		\vspace{-18pt}
		\caption{Training Accuracy vs. $\epsilon$}
		\label{fig:gowalla-noise-acc}
	\end{subfigure}
	\vspace{-10pt}
	\caption{Impact of applying adversarial noises and random noises to the parameters of MF-BPR on Pinterest and Gowalla.} \vspace{-5pt}
	\label{fig:noise}
\end{figure*}

Owing to its rationality and ease of optimization, BPR has been used in a wide variety of scenarios~\cite{ACF,JRL,CKBE,LambdaFM,Yu:2018:ACR,cao2017embedding} and plays an important role in optimizing recommender models. 
It is worth noting that the behavior of BPR can be interpreted as a classifier --- given a triplet $(u,i,j)$, it determines whether $(u,i)$ should have a higher score than $(u,j)$. Under this interpretation, a positive instance of $(u,i,j)$ means that $\hat{y}_{ui}$ should be larger than $\hat{y}_{uj}$ as much as possible to get the correct label of $+1$; and vice versa, a negative instance can be seen as having a label of $0$. 




\subsection{MF-BPR is Vulnerable to Adversarial Noises}
\label{ss:motivating-example}
Inspired by the findings of adversarial examples in image classification~\cite{AML_2014,AML_2015,moosavi2016universal}, we are particularly interested in exploring whether the similar phenomenon exists for BPR, since it can also be seen as a classification method with triplet $(u,i,j)$ as the input. 
Distinct from the image domain where adding small noises to an input image shall not change its visual content, the input to BPR is discrete ID features and changing an ID feature will change the semantics of the input. For example, if we change an input $(u,i,j)$ to $(u',i,j)$ by corrupting the user ID, the semantics of the triplet becomes totally different and the label may change. 
As such, existing methods that generate adversarial examples for an image classifier are inappropriate for BPR. 

Since it is irrational to add noises in the input layer, we instead consider exploring the robustness of BPR at a deeper level --- the parameters of the underlying recommender model. It is natural to assume that a robust model should be rather insensitive to small perturbations on its parameters; that is, only when large perturbations are enforced, the model behavior should be changed dramatically. To benchmark the perturbations needed, we use random perturbations as the baseline. If we can find a way to perturb the models parameters more effectively than random perturbations, \ie resulting in a much worse recommendation performance, it means that the model is not that robust and is vulnerable to certain perturbations.

\noindent\textbf{Settings}. Considering the dominant role of MF in recommendation, we choose MF as the recommender model and optimize it with BPR. We first train MF-BPR until convergence using SGD. We then compare the effect of adding random perturbations and adversarial perturbations to the embeddings of MF. For \textit{adversarial perturbations}, we define it as the perturbations that aim to maximize the objective function of BPR:
\begin{equation}\label{eq:adv}
	\Delta_{adv} = \arg\max_{\Delta,||\Delta||\leq \epsilon} L_{BPR}(\mathcal{D} | \hat{\Theta} + \Delta),
\end{equation}
where $\epsilon$ controls the magnitude of adversarial perturbations, $||\cdot||$ denotes the $L_2$ norm, and $\hat{\Theta}$ is a constant set denoting the current model parameters. As MF is a bilinear model and BPR objective function involves nonlinear operations, it is intractable to get exact maximization with respect to $\Delta$. Inspired by the fast gradient method proposed in Goodfellow \etal~\cite{AML_2015}, we approximate the objective function by linearizing it round $\Delta$. With this approximation and the max-norm constraint, we can obtain the optimal $\Delta$ as:
\begin{equation}
	\Delta_{adv} = \epsilon \frac{\Gamma}{||\Gamma||}\quad \text{where} \quad \Gamma = \frac{\partial L_{BPR}(\mathcal{D} | \hat{\Theta} + \Delta)}{\partial \Delta}.
\end{equation}
As the number of training instances in $\mathcal{D}$ is huge, we sample one unobserved item $j$ to pair with an observed interaction $(u,i)$. We then perform experiments on this reduced set of examples $\mathcal{D}'$ to verify the effect of adversarial perturbations.

\noindent\textbf{Results}. Figure \ref{fig:noise} shows the impact of applying adversarial and random perturbations to MF-BPR with different settings\footnote{Note that we enforce the max-norm constraint of $\epsilon$ on each embedding vector in $\textbf{P}$ and $\textbf{Q}$, rather than the whole matrix.} of $\epsilon$ on our Pinterest and Gowalla datasets (details see Section~\ref{ss:setting}). Specifically, we show the performance evaluated by NDCG@100 on the holdout testing set (Figure \ref{fig:noise}(a,c)) and the classification accuracy on the reduced training set $\mathcal{D}'$ (Figure \ref{fig:noise}(b,d)). 
The setting of $\epsilon=0$ means no perturbations are used, indicating the performance of the well-trained MF-BPR. We have two main observations. 
\begin{itemize}[leftmargin=*]
\item First, both datasets show that adding adversarial noises leads to a more significant performance drop than adding random noises. For example on Gowalla, when $\epsilon$ is set to 0.4, applying random perturbations decreases the testing NDCG by $1.6\%$, which is a very minor impact on recommendation; in contrast, applying adversarial perturbations decreases NDCG significantly by $21.2\%$ --- 13 times larger than that of random perturbations. 
\item Second, even though the adversarial perturbations are derived based on partial training instances $\mathcal{D}'$ only, it has a significant adverse effect on the recommendation performance. For example on Gowalla, when $\epsilon$ is set to 1, NDCG decreases by $55.4\%$, whereas the training accuracy decreases by $5.1\%$ only. Similar finding applies to the Pinterest dataset, where the drop of testing NDCG and training accuracy at $\epsilon=2$ are $57.8\%$ and $10.1\%$, respectively.
\end{itemize}
Our results indicate that MF-BPR is relatively robust to random noises, but it is rather vulnerable to certain perturbations that are purposefully designed. 
If a recommender model is robust and can predict user preference well, how can it be confused so much by perturbations at a small scale? The existence of such effective adversarial perturbations implies that the model learns a complicated function that overfits the training data and does not generalize well. This motivates us to develop new training methods for personalized ranking, which can lead to robust recommender models that are insensitive to such adversarial perturbations. 

 
\section{Proposed Methods}
\label{sec:method}
In this section, we first present APR, an adversarial learning framework for personalized ranking. We then derive a generic solver for APR based on SGD. Lastly, we present the AMF method, an instantiation of APR that uses MF as the recommender model. 

\subsection{Adversarial Personalized Ranking}
Our target is to design a new objective function such that by optimizing it, the recommender model is both suitable for personalized ranking and robust to adversarial perturbations. Due to the rationality of the BPR pairwise objective in personalized ranking, we choose it as the building block. To enhance the robustness, we enforce the model to perform well even when the adversarial perturbations (defined in Equation~(\ref{eq:adv})) are presented. To achieve this, we additionally optimize the model to minimize the BPR objective function with the perturbed parameters. 
Formally, we define the objective function of adversarial personalized ranking as follows:
\begin{equation}\label{eq:L_adv}
\begin{aligned}
&L_{APR}(\mathcal{D}|\Theta) = L_{BPR}(\mathcal{D} | \Theta) + \lambda L_{BPR}(\mathcal{D} | \Theta + \Delta_{adv}), \\
&\text{where}\quad \Delta_{adv} = \arg\max_{\Delta,||\Delta||\leq \epsilon} L_{BPR}(\mathcal{D} | \hat{\Theta} + \Delta),
\end{aligned}
\end{equation}
where $\Delta$ denotes the perturbations on model parameters, $\epsilon\geq 0$ controls the magnitude of the perturbations, and $\hat{\Theta}$ denotes the current model parameters. In this formulation, the adversarial term $L_{BPR}(\mathcal{D}|\Theta+\Delta_{adv})$ can be seen as regularizing the model by stabilizing the classification function in BPR. As such, we also call it as \textit{adversarial regularizer} and use $\lambda$ to control its strength. 
As the intermediate variable $\Delta$ maximizes the objective function to be minimized by $\Theta$, the training process of APR can be expressed as playing a minimax game:
\begin{equation}
\Theta^*, \Delta^* = \arg\min_{\Theta} \max_{\Delta, ||\Delta||\leq \epsilon} L_{BPR}(\mathcal{D}|\Theta) + \lambda L_{BPR}(\mathcal{D}|\Theta+\Delta),
\end{equation}
where the learning algorithm for model parameters $\Theta$ is the minimizing player, and the procedure for getting perturbations $\Delta$ acts as the maximizing player, which aims to identify the worst-case perturbations against the current model.
The two players alternately play the game until convergence. Since the focus of APR is to get a good recommender model, in practice we can determine when to stop the adversarial training by tracking how does the model perform on a validation set. 

We can see that, similar to BPR, our formulation of APR leads to a general learning framework which is model independent. As long as the underlying model $\hat{y}_{ui}(\Theta)$ is differentiable, it can be learned under our APR framework using backpropagation and gradient-based optimization algorithms. There are two hyper-parameters --- $\epsilon$ and $\lambda$ --- to be specified in APR in addition to the ones in BPR. 
In what follows, we propose a generic solution for APR based on SGD. 

\subsection{A Generic SGD Solver for APR}
Two optimization strategies are most widely used in recommendation --- coordinate descent (CD) and stochastic gradient descent (SGD). 
A typical example of CD is alternating least squares~\cite{fastMF}, which iterates through model parameters and updates one parameter at a time.
Note that CD is mostly used to optimize the pointwise regression loss on linear models~\cite{iCD}. When the optimization target involves nonlinearities, SGD becomes the default choice due to its ease in deriving the update strategy~\cite{CDAE,NCF}. 
Since APR involves nonlinear function in its objective function and it has a huge number of training instances (same as BPR), we optimize APR with SGD, which is easier to implement and is more efficient than CD. 

The idea of SGD is to randomly draw a training instance and update model parameters with respect to the single instance only. So we consider how to optimize model parameters with respect to a randomly sampled instance $(u,i,j)$. 

\textbf{Step 1. Constructing Adversarial Perturbations}. 
Given a training instance $(u,i,j)$, the problem of constructing adversarial perturbations $\Delta_{adv}$ can be formulated as maximizing
\begin{equation}\label{eq:adv_pert}
l_{adv}((u,i,j)|\Delta) = - \lambda\ln \sigma(\hat{y}_{ui}(\hat{\Theta}+\Delta) -\hat{y}_{uj}(\hat{\Theta}+\Delta)).
\end{equation}
Here $\hat{\Theta}$ is a constant set denoting current model parameters. As such, the $L_2$ regularizer for $\Theta$ is dropped since it is irrelevant to $\Delta$. 
However, for many models of interest such as the bilinear MF and multi-layer neural networks~\cite{NCF,CDAE},  it is difficult to get the exact optimal solution of $\Delta_{adv}$. Thus, we employ the fast gradient method proposed in Goodfellow \etal~\cite{AML_2015}, a common choice in adversarial training~\cite{miyato2016adversarial,ATforRE,AdversarialDropout}. The idea is to approximate the objective function around $\Delta$ as a linear function. To maximize the approximated linear function, we only need to move towards the gradient direction of the objective function with respect to $\Delta$, which can be derived as\footnote{Note the used derivative rules are: $\frac{\partial \ln x}{\partial x} = \frac{1}{x}, $ and $\frac{\partial \sigma(x)}{\partial x} = \sigma(x) (1-\sigma(x))$.}:
\begin{equation}
	\frac{\partial l_{adv}((u,i,j)|\Delta)}{\partial \Delta} = -\lambda (1 - \sigma(\hat{y}_{uij}(\hat{\Theta}+\Delta)))\frac{\partial \hat{y}_{uij}(\hat{\Theta} + \Delta) } {\partial \Delta},
\end{equation}
where $\hat{y}_{uij}(x) = \hat{y}_{ui}(x) -\hat{y}_{uj}(x)$ for short. With the max-norm constraint $||\Delta||\leq \epsilon$, we have the solution for $\Delta_{adv}$ as:
\begin{equation}\label{eq:delta_adv}
	\Delta_{adv} = \epsilon \frac{\Gamma}{||\Gamma||}\quad \text{where} \quad \Gamma = \frac{\partial l_{adv}((u,i,j)|\Delta)}{\partial \Delta}.
\end{equation}

\textbf{Step 2. Learning Model Parameters}. We now consider how to learn model parameters $\Theta$. The local objective function to minimize for a training instance $(u,i,j)$ is as follows:
\begin{equation}
\begin{aligned}
	l_{APR}((u,i,j)|\Theta) =& -\ln \sigma(\hat{y}_{ui}(\Theta)-\hat{y}_{uj}(\Theta)) + \lambda_{\Theta}||\Theta||^2 \\
	&- \lambda \ln \sigma(\hat{y}_{ui}(\Theta+\Delta_{adv}) -\hat{y}_{uj}(\Theta+\Delta_{adv})).
\end{aligned}
\end{equation}
In this problem, $\Delta_{adv}$ is a constant obtained from Equation (\ref{eq:delta_adv}). The derivative of the objective function with respect to $\Theta$ is as follows:
\begin{equation}\label{eq:l_APR}
\begin{aligned}
\frac{\partial l_{APR}((u,i,j)|\Theta)}{\partial \Theta} =& -(1-\sigma(\hat{y}_{uij}(\Theta)))\frac{\partial \hat{y}_{uij}(\Theta)}{\partial \Theta} + 2\lambda_{\Theta}\Theta \\
&- \lambda (1 - \sigma(\hat{y}_{uij}(\Theta+\Delta_{adv}))) \frac{\partial \hat{y}_{uij}(\Theta+\Delta_{adv})}{\partial\Theta}.
\end{aligned}
\end{equation}
Then we can obtain the SGD update rule for $\Theta$:
\begin{equation}\label{eq:Theta}
	\Theta = \Theta - \eta \frac{\partial l_{APR}((u,i,j)|\Theta)}{\partial \Theta},
\end{equation}
where $\eta$ denotes the learning rate. \\\vspace{-5pt}

\noindent To summarize the SGD solver for APR, we give the training process in Algorithm~\ref{alg:APR}. In each training step (line 3-5), we first randomly draw a instance $(u,i,j)$. We then execute the update rule for adversarial perturbations and model parameters in sequential order. \\\vspace{-5pt}

\noindent\textbf{Initialization}. It is worth mentioning that the model parameters $\Theta$ are initialized by optimizing BPR (line 1), rather than randomly initialized. This is because the adversarial perturbations are only reasonable and necessary to add when the model parameters start to overfit the data. When the model is underfitting, normal training process is sufficient to get better parameters. 
Besides pre-training with BPR, another feasible strategy is to dynamically adjust $\epsilon$ that controls the level of perturbations during training. For example, it is possible to learn $\epsilon$ based on a holdout validation set. We leave this exploration as future work, since we find that the current pre-training strategy with a constant $\epsilon$ works quite well.

\begin{algorithm}[t]
	\caption{SGD learning algorithm for APR.}
	\label{alg:APR}
	\KwIn{Training data $\mathcal{D}$, adversarial noise level $\epsilon$, adversarial regularizer $\lambda$, $L_2$ regularizer $\lambda_{\Theta}$, learning rate $\eta$;}
	\KwOut{Model parameters $\Theta$;}
	Initialize $\Theta$ from BPR \;
	\While{Stopping criteria is not met} {
		Randomly draw $(u,i,j)$ from $\mathcal{D}$ \;
		\tcp{Constructing adversarial perturbations}
		$\Delta_{adv} \leftarrow$ Equation (\ref{eq:delta_adv}) \;
		\tcp{Updating model parameters}
		$\Theta \leftarrow $ Equation (\ref{eq:Theta}) \;
	}
	\Return{$\Theta$}
\end{algorithm}

\subsection{Adversarial Matrix Factorization}
\begin{figure}[t]
	\centering
	\includegraphics[width=0.48\textwidth]{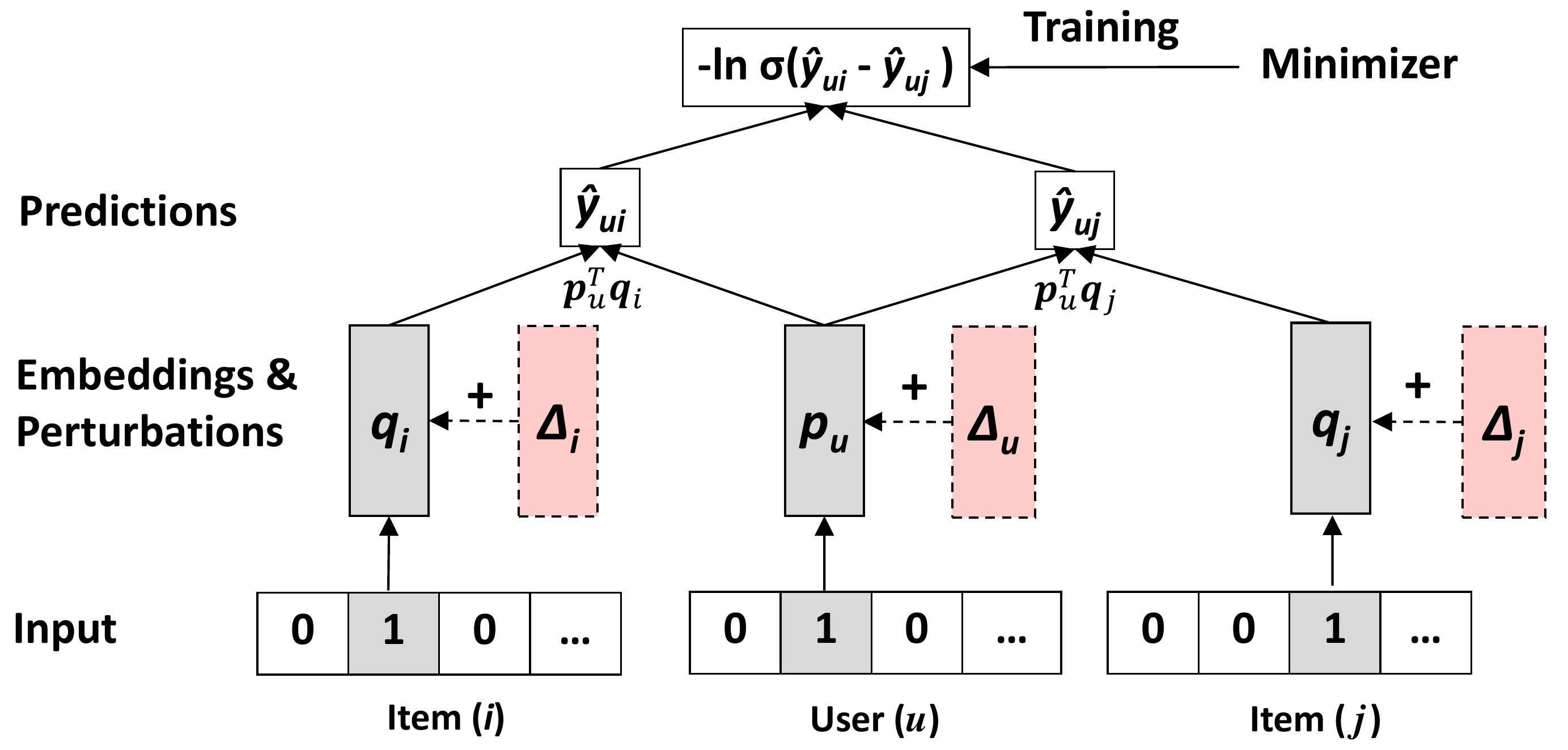} \vspace{-15pt}
	\caption{Illustration of our AMF method. The perturbations $\Delta$ are enforced on each embedding vector of user and item. }
	\vspace{-10pt}
	\label{fig:AMF}
\end{figure}

To demonstrate how the APR works, we now provide a specific recommender solution based on MF, a basic yet very effective model in recommendation. The solution is simple and straightforward --- we first train MF with BPR, and then further optimize it under our APR framework. We term the method as \textit{Adversarial Matrix Factorization} (AMF). Figure \ref{fig:AMF} illustrates our AMF method. Since the parameters of MF are embedding vectors for users and items, we apply adversarial perturbations on the embedding vector. Given a $(u,i)$ pair, the predictive model with perturbations is defined as:
\begin{equation}
	\hat{y}_{ui}(\Theta + \Delta) = (\textbf{p}_u + \Delta_{u})^T (\textbf{q}_i + \Delta_{i}),
\end{equation}
where $\Delta_u\in\mathbb{R}^K$ and $\Delta_i\in\mathbb{R}^K$ denote the perturbation vector for user $u$ and item $i$, respectively. Note that the max-norm constraint $||\Delta||\leq \epsilon$ is enforced on the level of perturbation vector. To apply Algorithm \ref{alg:APR} in AMF, we simply need to materialize Equation (\ref{eq:delta_adv}) and (\ref{eq:Theta}). For Equation (\ref{eq:delta_adv}), we give the key derivatives as: 
\begin{equation} \small
\begin{aligned}
	& \frac{\partial \hat{y}_{uij}(\hat{\Theta} + \Delta) } {\partial \Delta_u} = \textbf{q}_i + \Delta_{i} - \textbf{q}_j - \Delta_{j}, \\
	& \frac{\partial \hat{y}_{uij}(\hat{\Theta} + \Delta) } {\partial \Delta_i} = \textbf{p}_u + \Delta_u, \quad  \frac{\partial \hat{y}_{uij}(\hat{\Theta} + \Delta) } {\partial \Delta_j} = -\textbf{p}_u - \Delta_u.
\end{aligned}
\end{equation}
To implement Equation (\ref{eq:Theta}), we give the key derivatives as follows:
\begin{equation}
\begin{aligned} \small
	&\frac{\partial \hat{y}_{uij}(\Theta)}{\partial \textbf{p}_u} = \textbf{q}_i - \textbf{q}_j, \quad \frac{\partial \hat{y}_{uij}(\Theta)}{\partial \textbf{q}_i} = \textbf{p}_u, \quad \frac{\partial \hat{y}_{uij}(\Theta)}{\partial \textbf{q}_j} = -\textbf{p}_u, \\
	&\frac{\partial \hat{y}_{uij}(\Theta+\Delta_{adv})}{\partial \textbf{p}_u} = \textbf{q}_i + \Delta_{i} - \textbf{q}_j - \Delta_{j}, \\
	& \frac{\partial \hat{y}_{uij}(\Theta+\Delta_{adv})}{\partial \textbf{q}_i} = \textbf{p}_u + \Delta_u, \quad \frac{\partial \hat{y}_{uij}(\Theta+\Delta_{adv})}{\partial \textbf{q}_j} = -\textbf{p}_u - \Delta_u.
\end{aligned}
\end{equation}

\subsubsection{Mini-batch Training for AMF} Modern computing units such as CPU and GPU usually provide speedups for matrix-wise float operations. To leverage such speedups in learning complex models, a common strategy is to perform SGD in a mini-batch manner, \ie updating model parameters on a set of training instances rather than one instance only. In fact, many machine learning methods implemented in modern tools such as TensorFlow and Theano apply mini-batch optimizers. 
Since AMF plays a minimax game and has two coupled procedures, there are several ways to perform mini-batch training. 
Below we detail how we perform mini-batch training for AMF. 

First, given the mini-batch size $S$, we randomly draw $S$ training instances from $\mathcal{D}$ and term the mini-batch as $\mathcal{D}'$. We then construct adversarial perturbations by maximizing the adversarial regularizer over the mini-batch:
\begin{equation}
	L_{adv}(\mathcal{D}'|\Delta) = \sum_{(u,i,j)\in\mathcal{D}'} l_{adv}((u,i,j)|\Delta),
\end{equation}
where $l_{adv}((u,i,j)|\Delta)$ has been defined in Equation (\ref{eq:adv_pert}). For each user and item\footnote{Note that the item includes both positive item $i$ and negative item $j$. It is possible that a positive $i$ occurs in another instance as a negative item, and vice versa. This needs to be taken into account to avoid mistake.} that occurred in $\mathcal{D}'$, we compute its perturbed vector by enforcing the max-norm constraint on $\frac{\partial L_{adv}(\mathcal{D}'|\Delta)}{\partial \Delta}$. 

Next, we update the model parameters based on the mini-batch $\mathcal{D}'$. The APR objective function over the mini-batch is given as:
\begin{equation}
	L_{APR}(\mathcal{D}'|\Theta) = \sum_{(u,i,j)\in\mathcal{D}'} l_{APR}((u,i,j)|\Theta),
\end{equation}
where $l_{APR}((u,i,j)|\Theta)$ has been defined in Equation (\ref{eq:l_APR}). Similarly, for each user and item that occurred in $\mathcal{D}'$, we perform a SGD update as 
$\Theta = \Theta - \eta \frac{\partial L_{APR}(\mathcal{D}'|\Theta)}{\partial \Theta}$. We iterate the above two steps until AMF reaches a convergence state or the validation performance starts to degrade. 


\section{Experiments}
\label{sec:experiment}
As the key contribution of this work is to develop a new adversarial learning method APR for personalized ranking, we aim to answer the following research questions via experiments. 
\begin{itemize}
	\item[\textbf{RQ1}] How is the effect of adversarial learning? Can AMF improve over MF-BPR by performing adversarial learning? 
	\item[\textbf{RQ2}] How does AMF perform compared with state-of-the-art item recommendation methods?
	\item[\textbf{RQ3}] How do the hyper-parameters $\epsilon$ and $\lambda$ affect the performance and how to choose optimal values? 
\end{itemize}
Next, we first describe the experimental settings. We then report results by answering the above research questions in turn. 

\subsection{Experimental Settings}
\label{ss:setting}
\subsubsection{Datasets} We experiment with three publicly available datasets. Table \ref{tab:dataset} summarizes the statistics of the datasets (after all pre-processing steps). These three million-size scale datasets represent different item recommendation scenarios for business, image, and location check-in. 
\begin{table}[h]
	\begin{center}
		\caption{\textbf{Statistics of the experimented datasets.}}
		\vspace{-10pt}
		\small
		\label{tab:dataset}
		\begin{tabular}{ | l | c | c | c | c | }
			\hline
			\textbf{Dataset} & \textbf{Interaction\#} & \textbf{Item\#} & \textbf{User\#}  & \textbf{Sparsity} \\ \hline
			Yelp	& 730,790 & 25,815 &  25,677 & 99.89\% \\ \hline
			Pinterest	& 1,500,809	& 9,916 & 55,187 & 99.73\% \\ \hline
			Gowalla & 1,249,703 & 52,400 & 54,156 & 99.96\% \\ \hline
		\end{tabular}
	\end{center}
\end{table}

\textbf{1. Yelp}\footnote{Downloaded from: \url{https://github.com/hexiangnan/sigir16-eals}}. This is the Yelp Challenge data of user ratings on businesses. We use the filtered subset created by \cite{fastMF} for evaluating item recommendation. We find that a user may rate an item multiple times at different timestamps. Since a recommender system typically aims to recommend items that a user did not consume before, we further merge repetitive ratings to the earliest one. This can also avoid a testing interaction appearing in the training set. 

\textbf{2. Pinterest}\footnote{Downloaded from: \url{https://github.com/hexiangnan/neural_collaborative_filtering}}. This implicit feedback dataset was originally constructed by \cite{DBLP:conf/iccv/GengZBC15} for content-based image recommendation. We use the filtered subset created by \cite{NCF} for evaluating collaborative recommendation on images. Since no repetitive interactions are found, we use the downloaded dataset as it is. 

\textbf{3. Gowalla}\footnote{Downloaded from: \url{http://dawenl.github.io/data/gowalla_pro.zip}}. This is the check-in dataset constructed by \cite{ExpoMF} for item recommendation. Each interaction represents a user's check-in behavior on a venue in Gowalla, a location-based social network. Same as the setting of Yelp, we merge repetitive check-ins to the earliest check-in. We then filter out items that have less than 10 interactions and users that have less than 2 interactions.

\subsubsection{Evaluation Protocol} We employ the standard \textit{leave-one-out} protocol, which has been widely used in item recommendation evaluation~\cite{BPR,fastMF,iCD}. Specifically, for each user in Yelp and Gowalla, we hold out the latest interaction as the testing set and train a model on the remaining interactions. As the Pinterest data has no timestamp information, we randomly hold out an interaction for each user to form the testing set. 

After a model is trained, we generate the personalized ranking list for a user by ranking all items that are not interacted by the user in the training set. 
To study the performance of top-$K$ recommendation, we truncate the ranking list at position $K$; the default setting of $K$ is 100 without special mention.  
We then evaluate the ranking list using \textit{Hit Ratio}~(HR) and \textit{Normalized Discounted Cumulative Gain}~(NDCG). HR is a recall-based metric, measuring whether the testing item is in the top-$K$ list. While NDCG is position-sensitive, which assigns higher score to hits at higher positions.
For both metrics, larger values indicate better performance. We report the average score for all users, and perform one-sample paired t-test to judge the statistical significance where necessary.

\begin{figure}[t]
	\centering
	\begin{subfigure}[b]{0.235\textwidth}
		\centering
		\includegraphics[width=\textwidth]{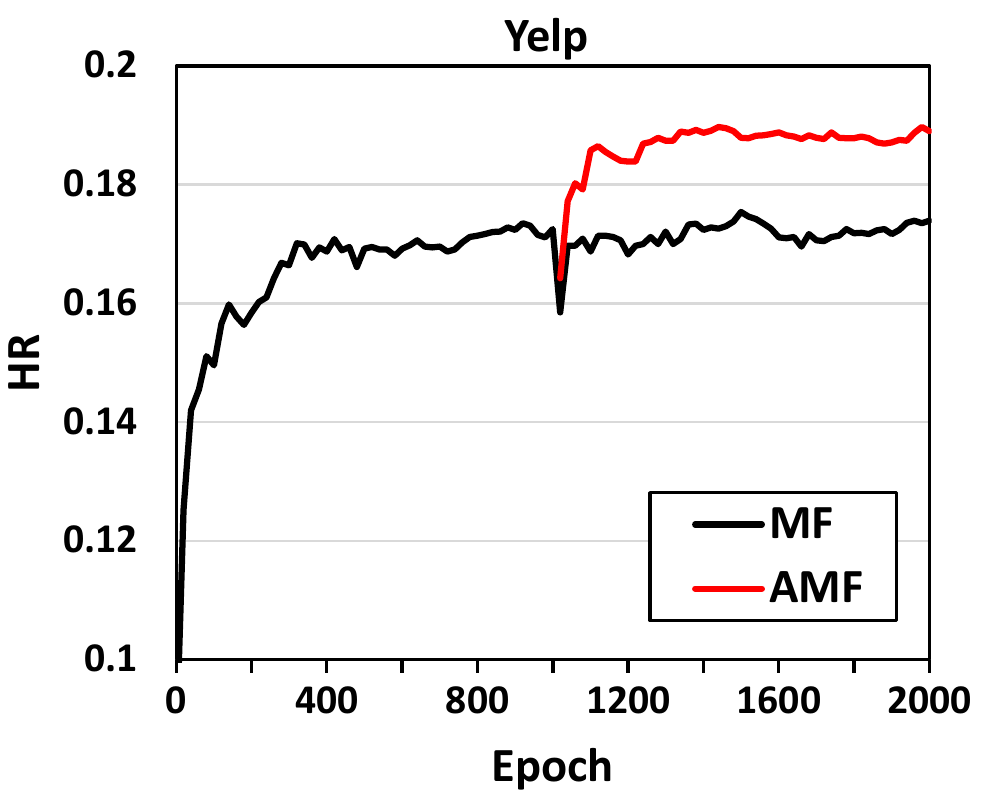}
		\vspace{-20pt}
	\end{subfigure} 
	\begin{subfigure}[b]{0.235\textwidth}
		\centering
		\includegraphics[width=\textwidth]{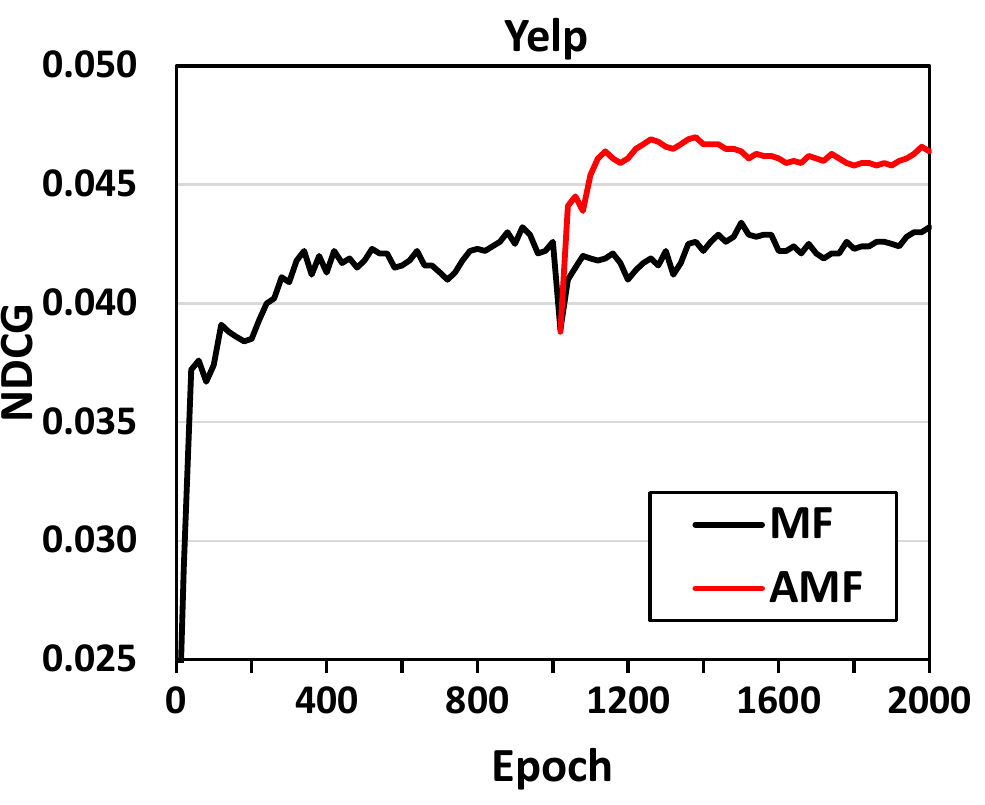}
		\vspace{-20pt}
	\end{subfigure} 
	\caption{Training curves of MF-BPR and AMF on Yelp.} \vspace{-15pt}
	\label{fig:yelp-epoch}
\end{figure}

\subsubsection{Baselines} We compare with the following methods:

- \textbf{ItemPop}. This method ranks items based on their popularity, evidenced by the number of interactions in the training set. This is a non-personalized method to benchmark the performance of personalized recommendation. 

- \textbf{MF-BPR}~\cite{BPR}. This method optimizes MF with the BPR objective function. It is a highly competitive approach for item recommendation. We tuned the learning rate and the coefficient for $L_2$ regularization. 

- \textbf{CDAE}~\cite{CDAE}. This method extends the \textit{Denoising Auto-Encoder} for item recommendation. It has been shown to be able to generalize several latent factor models. We used the original implementation released by the authors\footnote{\url{https://github.com/jasonyaw/CDAE}}, and tuned the hyperparameters in the same way as reported in their paper, including the loss function, corruption level, $L_2$ regularization and learning rate. 

\begin{figure}[t]
	\centering
	\begin{subfigure}[b]{0.235\textwidth}
		\centering
		\includegraphics[width=\textwidth]{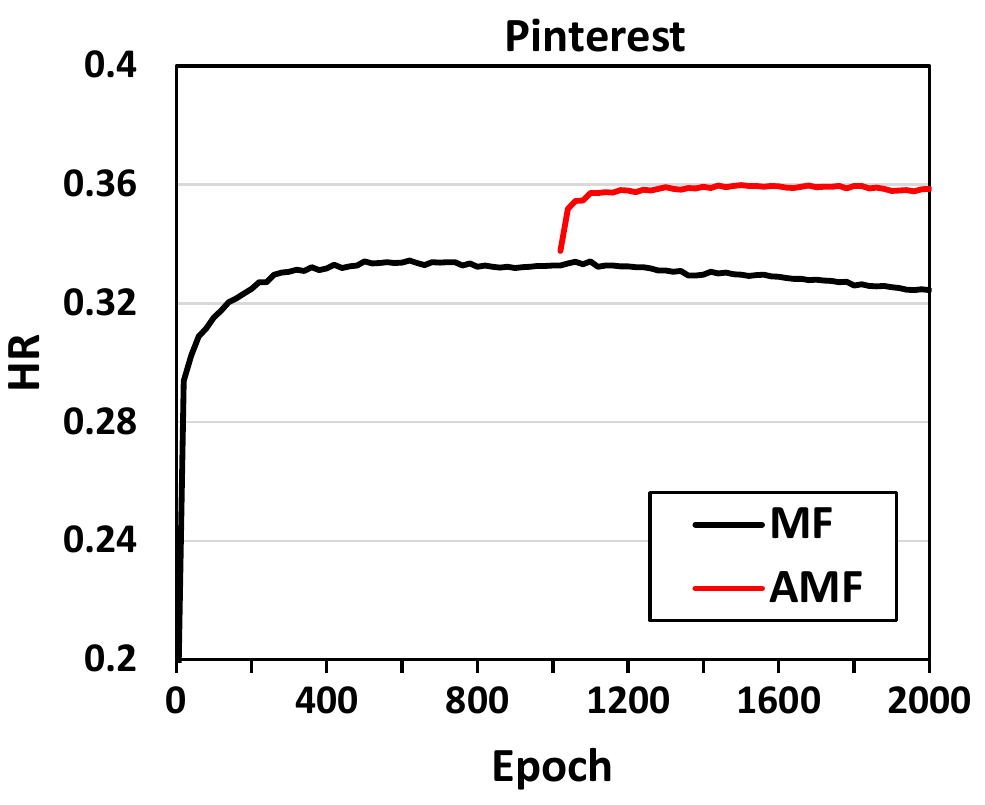}
		\vspace{-20pt}
	\end{subfigure} 
	\begin{subfigure}[b]{0.235\textwidth}
		\centering
		\includegraphics[width=\textwidth]{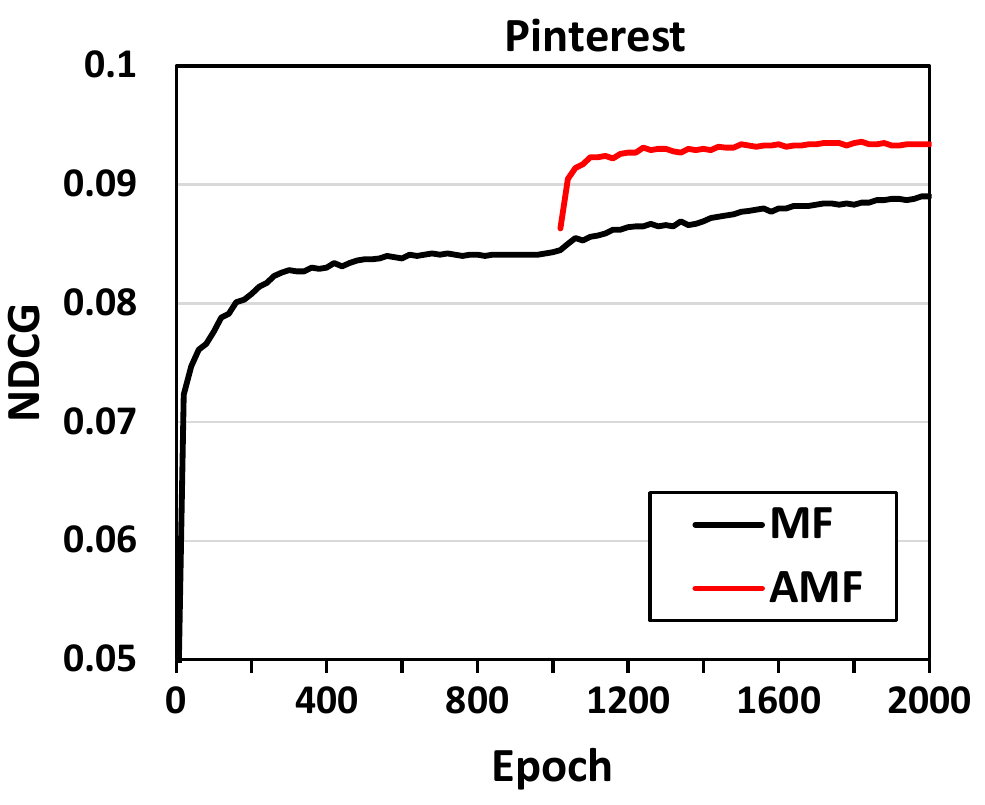}
		\vspace{-20pt}
	\end{subfigure} 
	\caption{Training curves of MF-BPR and AMF on Pinterest.} \vspace{-10pt}
	\label{fig:pinterest-epoch}
\end{figure}
\begin{figure}[t]
	\centering
	\begin{subfigure}[b]{0.235\textwidth}
		\centering
		\includegraphics[width=\textwidth]{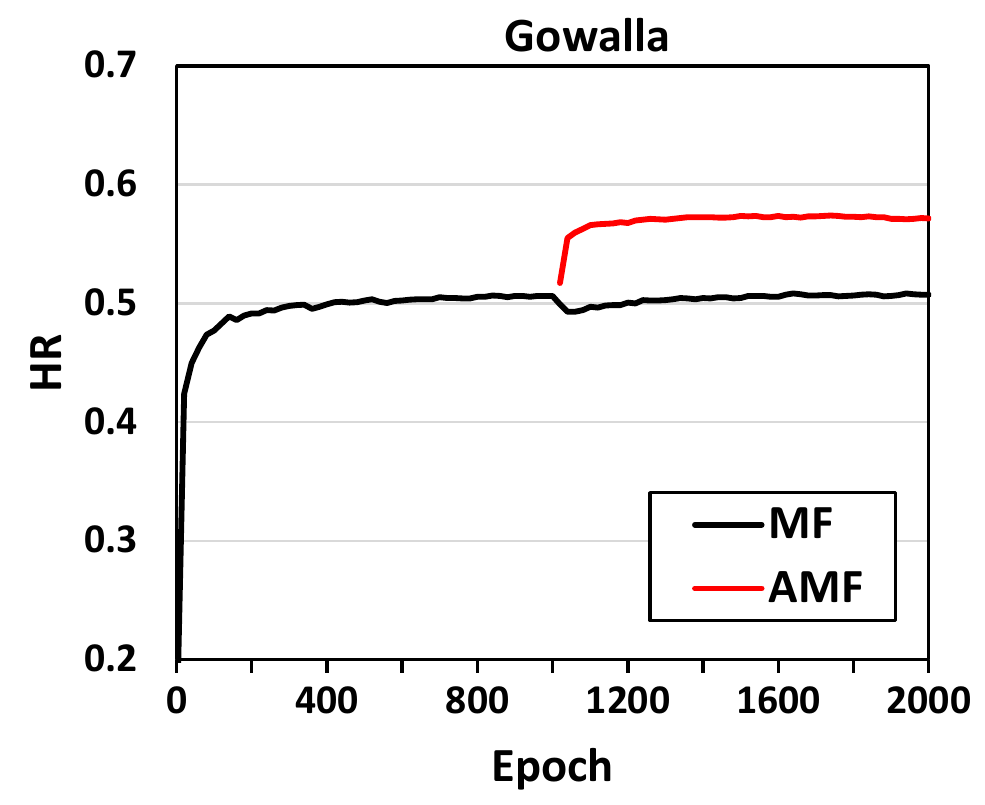}
		\vspace{-20pt}
	\end{subfigure} 
	\begin{subfigure}[b]{0.235\textwidth}
		\centering
		\includegraphics[width=\textwidth]{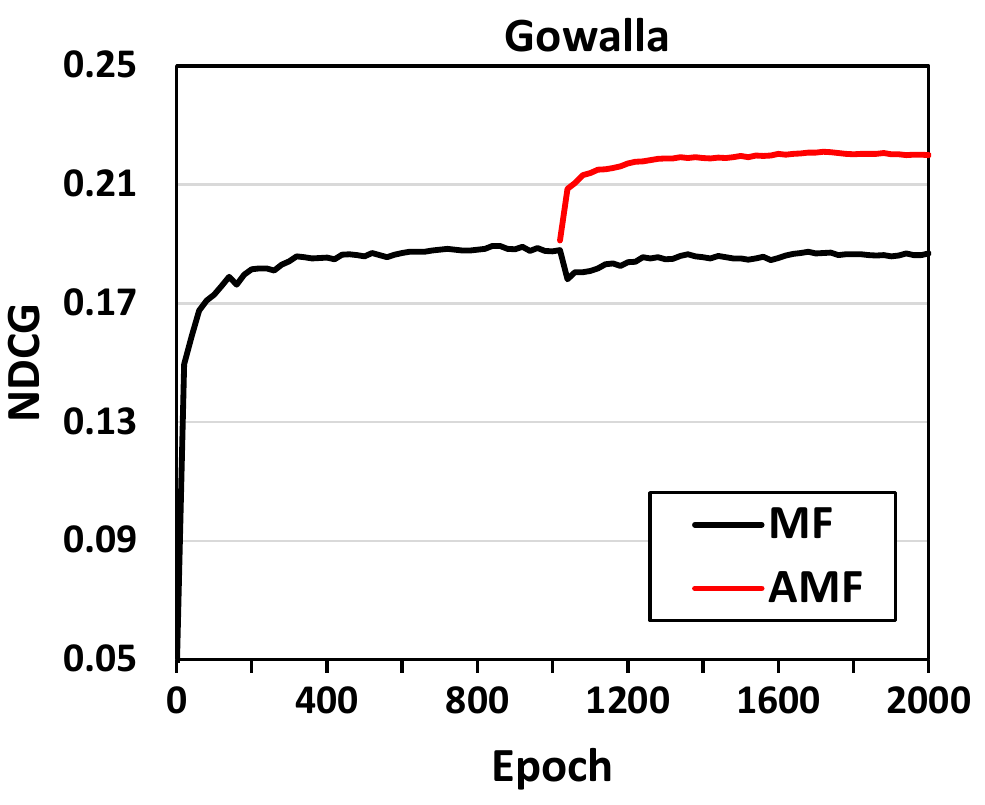}
		\vspace{-20pt}
	\end{subfigure} 
	\caption{Training curves of MF-BPR and AMF on Gowalla.} \vspace{-10pt}
	\label{fig:gowalla-epoch}
\end{figure}

- \textbf{NeuMF}~\cite{NCF}. \textit{Neural Matrix Factorization} is the state-of-the-art item recommendation method. It combines MF and multi-layer perceptrons (MLP) to learn the user-item interaction function. 
As suggested in the paper, we pre-trained the model with MF, and tuned the depth and $L_2$ regularizer for the hidden layers. 

- \textbf{IRGAN}~\cite{IRGAN}. This method combines two types of models via adversarial training, a generative model that generates items for a user and a discriminative model that determines whether the instance is from real data or generated. We used the implementation released by the authors\footnote{\url{https://github.com/geek-ai/irgan}}. We followed the setting of the paper that pre-trains the generator with LambdaFM~\cite{LambdaFM}. We tuned the learning rate and number of epochs for generator and discriminator separately, which we found to have a large impact on its performance. Further tuning of the sampling temperature did not improve the results, so we used their default settings. \vspace{+5pt}

\begin{figure*}[t]
	\centering
	\begin{subfigure}[b]{0.3\textwidth}
		\centering
		\includegraphics[width=\textwidth]{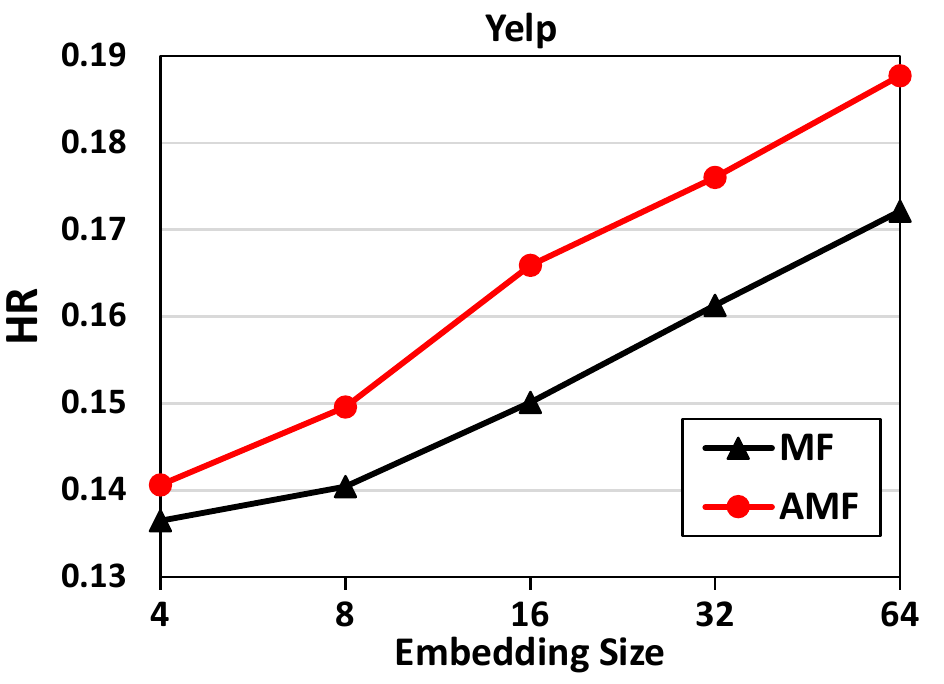}
		\vspace{-18pt}
		\label{fig:yelp-embedding-hr}
	\end{subfigure} \hspace{+5pt}
	\begin{subfigure}[b]{0.3\textwidth}
		\centering
		\includegraphics[width=\textwidth]{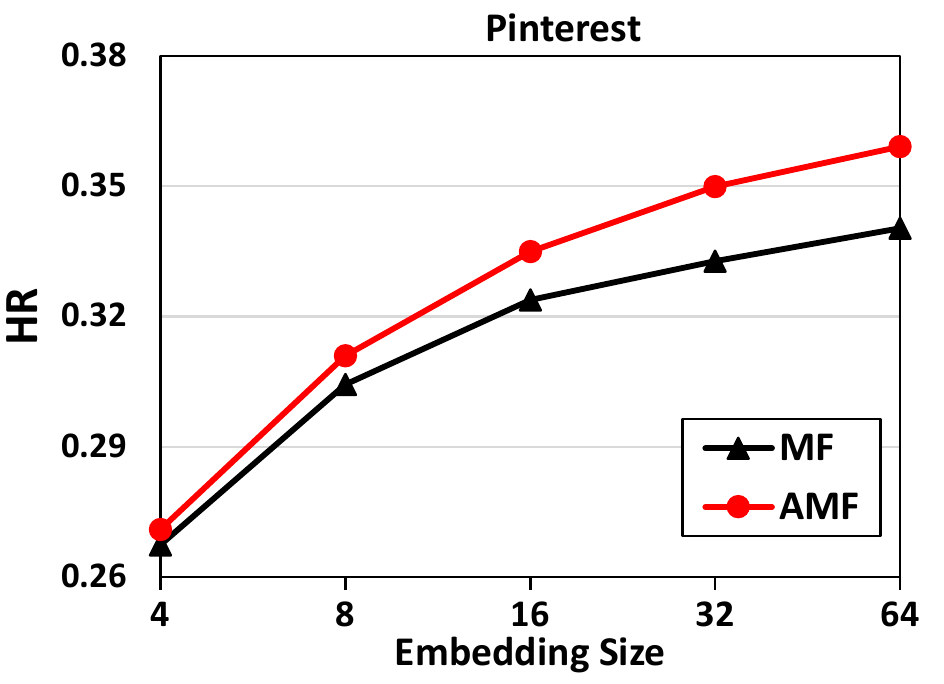}
		\vspace{-18pt}
		\label{fig:pinterest-embedding-hr}
	\end{subfigure} \hspace{+5pt}
	\begin{subfigure}[b]{0.3\textwidth}
		\centering
		\includegraphics[width=\textwidth]{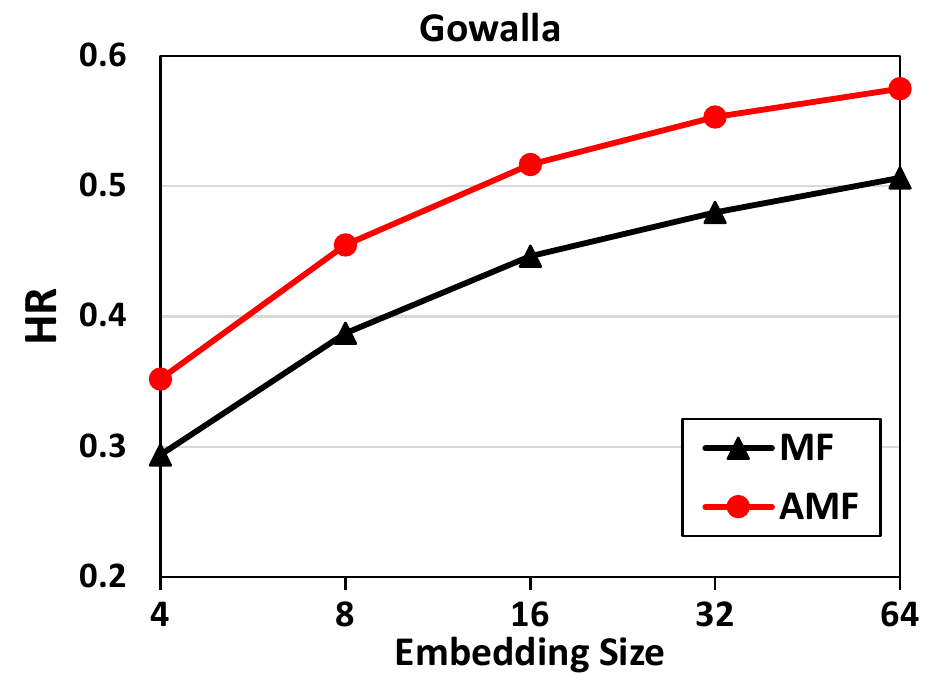}
		\vspace{-18pt}
		\label{fig:gowalla-embedding-hr}
	\end{subfigure} \hspace{+5pt}
	\caption{Performance comparison of HR between MF-BPR and AMF with respect to different embedding sizes.} \vspace{-10pt}
	\label{fig:embedding_hr}
\end{figure*}

\noindent This set of baselines stands for the state-of-the-art performance for the item recommendation task. In particular, CDAE and NeuMF are the recently proposed neural recommender models which have shown significant improvements over conventional shallow methods like MF and FISM~\cite{FISM}. IRGAN takes advantage of generative adversarial networks~\cite{GAN} and shows good performance on several IR tasks including recommendation in their paper.

\subsubsection{Implementation and Parameter Settings} Our implementation is based on TensorFlow, which is available at: \url{https://github.com/hexiangnan/adversarial_personalized_ranking}.
To tune the hyper-parameters, we randomly holdout one interaction for each user from the training interactions as the validation set, and we choose the optimal hyperparameters based on NDCG@100. 
For a fair comparison, all models are set with an embedding size of 64 and optimized using the mini-batch Adagrad~\cite{Adagrad} with a batch size of 512; moreover, the learning rate is tuned in $[0.005,0.01,0.05]$.
For AMF, we tune $\epsilon$ in $[0.001,
0.005,0.01,...,1,5]$ and $\lambda$ in $[0.001,0,01,...,1000]$. With MF-BPR as pre-training, AMF achieves good performance when $\epsilon=0.5$ and $\lambda=1$ on all datasets. As such, without special mention, we report the performance of AMF on this specific setting. 

\subsection{Effect of Adversarial Learning (RQ1)}
To validate the effect of adversarial learning, we first train MF with BPR for $1,000$ epochs (mostly converged), where each epoch is defined as training the number of instances the same as the size as the training set. We then continue training MF with APR, \ie our proposed AMF method; as a comparison, we further train MF with BPR to be consistent with APR.

\noindent\textbf{1. Training Process}. Figure \ref{fig:yelp-epoch} to \ref{fig:gowalla-epoch} show the performance of MF and AMF evaluated per 20 epochs on the three datasets. 
We can see that all figures show the same trend --- after $1,000$ epochs, further training MF with APR leads to a significant improvement, whereas further training MF with BPR has little improvements. For example, on Yelp (Figure~\ref{fig:yelp-epoch}) the best HR and NDCG of MF-BPR are $0.1721$ and $0.0420$, respectively, which are improved to $0.1881$ and $0.0470$ by training with APR. This roughly $10\%$ relative improvement is very remarkable in recommendation, especially considering that the underlying recommender model remains the same and we only change the way of training it. 

On Gowalla~(Figure~\ref{fig:gowalla-epoch}), the improvements are even larger --- $13.5\%$ and $16.8\%$ in terms of HR and NDCG, respectively. On Pinterest (Figure~\ref{fig:pinterest-epoch}), we notice that HR and NDCG of MF exhibit different trends, where after $1,000$ epochs HR starts to decrease while NDCG keeps increasing. This is understandable, since HR and NDCG measure different aspects of a ranking list --- NDCG is position-sensitive by assigning higher rewards to hits at higher positions while HR is not. Moreover, this points to the strength of BPR in ranking top items, owing to its pairwise objective. This observation is consistent with \cite{fastMF}'s finding in evaluating top-K recommendation. 

\noindent \textbf{2. Improvements vs. Model Size}. 
Furthermore, we investigate whether the advantages of adversarial learning apply to models of different sizes. Figure~\ref{fig:embedding_hr} show the performance of MF-BPR and AMF with respect to different embedding sizes. Note that we show the results of HR only due to space limitation, and the figures of NDCG admit the same findings. First, we can see a clear trend that the performance of both methods increase with a larger embedding size. This indicates that a larger model is beneficial to top-K recommendation due to the increased modeling capability. Second, we observe that AMF demonstrates consistent improvements over MF on models of all embedding sizes. 
Notably, AMF with an embedding size of 32 even performs better than MF with a larger embedding size of 64 on all datasets. This further verifies the positive effect of adversarial learning in our APR method. 

Lastly, it is worth noting that the improvements of AMF are less significant when the embedding size is small, compared to the setting of large embedding size. This implies that when a model is small and has limited capability, its robustness is not a serious issue. While for large models that are easy to overfit the training data, it is crucial to increase a model's robustness by learning with adversarial perturbations, which in turn can increase its generalization performance. We believe that this insight is particularly useful for the recommendation task, which typically involves a large space of input features (\eg user ID, item ID, and other attributed and contextual variables). Given such a large feature space, even a shallow embedding model like \textit{Factorization Machine}~\cite{FM} will have a large number of parameters, not to mention the more expressive deep neural networks such as \textit{Neural Factorization Machine}~\cite{NFM} and \textit{Deep Crossing}~\cite{DeepCrossing}. This work introduces adversarial learning to address the ranking task, providing a new means to increase the generalization ability of large models and having the potential to improve a wide range of models. \vspace{-10pt}
\begin{table}[h]
	\begin{center}
		\caption{The impact of applying adversarial perturbations to the MF model trained by BPR and APR, respectively. The number shows the relative decrease in NDCG.} \vspace{-10pt}
		\label{tab:noise}
		\small
		\begin{tabular}{| l | c | c | c | c | c | c |} \hline
			& \multicolumn{2}{c|}{$\epsilon=0.5$} & \multicolumn{2}{c|}{$\epsilon=1.0$} & \multicolumn{2}{c|}{$\epsilon=2.0$} \\ \hline
			\textbf{Dataset} & \textbf{BPR} & \textbf{APR} & \textbf{BPR} & \textbf{APR} & \textbf{BPR} & \textbf{APR} \\ \hline
			Yelp &-22.1$\%$ &-4.7$\%$ &-42.7$\%$ &-12.5$\%$ &-63.8$\%$ &-31.0$\%$ \\ \hline
			Pinterest & -9.5$\%$ &-2.6$\%$ &-25.1$\%$ &-7.2$\%$ &-55.7$\%$ &-23.4$\%$ \\ \hline
			Gowalla & -26.3$\%$ &-2.9$\%$ &-53.0$\%$ &-13.2$\%$ &-78.0$\%$ &-29.2$\%$ \\ \hline
		\end{tabular}
	\end{center}
\end{table}\vspace{-5pt}

\begin{table*}[t]
	\begin{center}
		\caption{Top-$K$ recommendation performance at $K=50$ and $K=100$. The best result of each setting is highlighted in bold font. $*$ indicates that the improvement of the best result is statistically significant for $p<0.01$ compared against all other methods. The last column ``RI'' indicates the relative improvement of AMF over the corresponding baseline on average.} \vspace{-10pt}
		\label{tab:main}
		\small
		\begin{tabular}{ l | c | c | c | c | c | c | c | c | c | c | c | c | c } \hline
			& \multicolumn{2}{c|}{\textbf{Yelp, HR}} & \multicolumn{2}{c|}{\textbf{Yelp, NDCG}} & \multicolumn{2}{c|}{\textbf{Pinterest, HR}} & \multicolumn{2}{c|}{\textbf{Pinterest, NDCG}} & \multicolumn{2}{c|}{\textbf{Gowalla, HR}} & \multicolumn{2}{c|}{\textbf{Gowalla, NDCG}} & \textbf{RI} \\ \hline
			& \textbf{K=50} & \textbf{K=100} & \textbf{K=50} & \textbf{K=100}& \textbf{K=50} & \textbf{K=100} & \textbf{K=50} & \textbf{K=100}& \textbf{K=50} & \textbf{K=100} & \textbf{K=50} & \textbf{K=100} & \\ \hline
			ItemPop & 0.0405 & 0.0742 & 0.0114 & 0.0169 & 0.0294 & 0.0485 & 0.0085 & 0.0116 & 0.1183 & 0.1560 & 0.0367 & 0.0428 & +416\% \\\hline
			MF-BPR & 0.1053 & 0.1721 & 0.0312 & 0.0420 & 0.2226 & 0.3403 & 0.0696 & 0.0886 & 0.4061 & 0.5072 & 0.1714 & 0.1878 & +11.2\% \\\hline
			CDAE~\cite{CDAE} & 0.1041 & 0.1733 & 0.0293 & 0.0405 & 0.2254 & 0.3495 & 0.0672 & 0.0873 & 0.4435 & 0.5483 & 0.1837 & 0.2007 & +9.5\% \\\hline
			IRGAN~\cite{IRGAN} & 0.1119 & 0.1765 & \textbf{0.0361}$^*$ & \textbf{0.0465}$^*$ & 0.2254 & 0.3363 & 0.0724 & 0.0904 & 0.4157 & 0.518 & 0.1853 & 0.2019 & +5.9\% \\\hline
			NeuMF~\cite{NCF} & 0.1135 & 0.1817 & 0.0335 & 0.0445 & 0.2342 & 0.3526 & 0.0734 & 0.0925 & 0.4558 & 0.5642 & 0.1962 & 0.2138 & +2.9\% \\\hline	
			AMF & \textbf{0.1176}$^*$ & \textbf{0.1885}$^*$ & 0.0350 & \textbf{0.0465}$^*$ & \textbf{0.2375}$^*$ & \textbf{0.3595}$^*$ & \textbf{0.0741}$^*$ & \textbf{0.0938}$^*$ & \textbf{0.4693}$^*$ & \textbf{0.5763}$^*$ & \textbf{0.2039}$^*$ & \textbf{0.2212}$^*$ & - \\\hline
		\end{tabular} \vspace{-5pt}
	\end{center}
\end{table*}

\noindent \textbf{3. Robustness of AMF}. We retrospect our motivating example in Section~\ref{ss:motivating-example} to investigate the robustness of a model trained by APR. Table~\ref{tab:noise} shows the impact of applying adversarial perturbations to the MF model trained by BPR and APR, respectively. 

We can see that by training MF with APR, the model becomes less sensitive to adversarial perturbations compared to that trained with BPR. For example, on Gowalla, adding adversarial perturbations at a noise level of $0.5$ to MF-BPR decreases NDCG by $26.3\%$, while the number is only $2.9\%$ for AMF. These results verify that our AMF is rather robust to adversarial perturbations, an important property that indicates good generalization ability of a model.

\begin{figure}[t]
	\centering
	\begin{subfigure}[b]{0.235\textwidth}
		\centering
		\includegraphics[width=\textwidth]{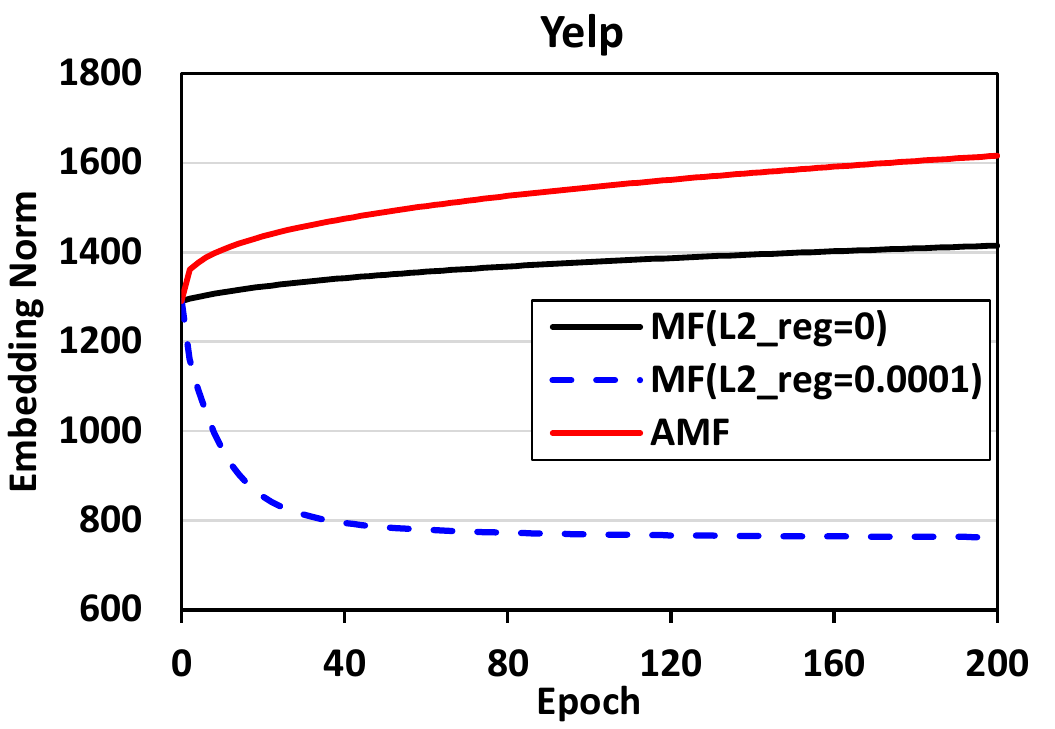}
		\vspace{-18pt}
	\end{subfigure} 
	\begin{subfigure}[b]{0.235\textwidth}
		\centering
		\includegraphics[width=\textwidth]{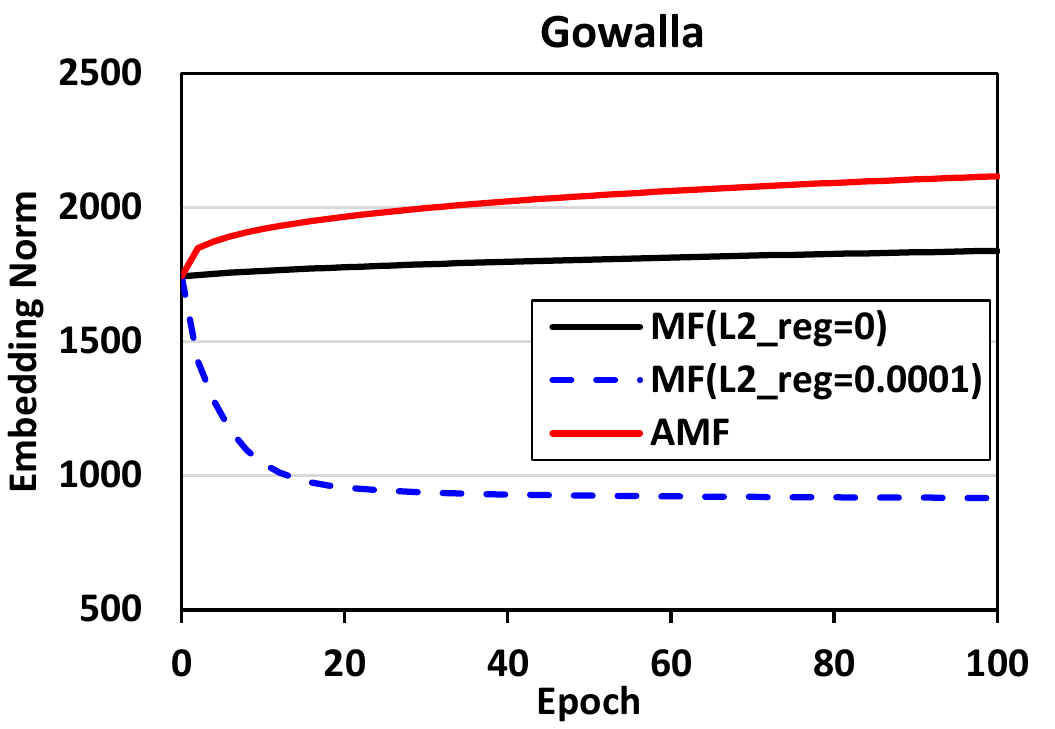}
		\vspace{-18pt}
	\end{subfigure} 
	\caption{The norm of embedding matrices of MF and AMF at each training epoch on Yelp and Gowalla.} \vspace{-15pt}
	\label{fig:norm-epoch}
\end{figure}

\noindent \textbf{4. Adversarial Regularization vs. $L_2$ Regularization}. The reason that APR improves over BPR is because of the adversarial regularizer. To be clear about its effect on parameter learning, we perform some micro-level analysis on model parameters. Figure \ref{fig:norm-epoch} shows the norm of embedding matrices (\ie $||\textbf{P}||^2+||\textbf{Q}||^2$) of MF and AMF in each epoch on Yelp and Gowalla. As a comparison, we also show the effect of $L_2$ regularization, a popular technique in recommendation to prevent overfitting.

Interestingly, we find that adding adversarial regularization increases the embedding norm. This is reasonable, since we constrain the adversarial perturbations in APR to have a fixed norm, thus increasing the embedding norm is helpful to reduce the impact of perturbations. Nevertheless, simply increasing the norm by scaling up the parameters is a trivial solution, which will not improve a model's generalization performance. This provides evidence that our proposed learning algorithm indeed updates parameters in a rather meaningful way towards enhancing the model's robustness. 
In contrast, adding $L_2$ regularization decreases the embedding norm to combat overfitting. 
Based on these, we conclude that adversarial regularization improves a model's generalization in a different but more effective way from the conventional $L_2$ regularization. 


\subsection{Performance Comparison (RQ2)}

We now compare our AMF with baselines. Table \ref{tab:main} shows the results of top-K recommendation with $K$ setting to 50 and 100. Note that we do not report results at smaller $K$, because our protocol ranks all items which makes the results at smaller $K$ exhibit large variances. More importantly, evaluating at a larger $K$ is more instructive for practitioners\footnote{Practical recommender systems typically have two stages~\cite{wang2018path}, 1) \textit{candidate selection} that selects hundreds of items that might be of interest to a user, and 2) \textit{ranking} that re-ranks the candidates to show top a few results. The first stage typically relies on collaborative filtering (CF) with the objective of a high recall. Thus, it is more instructive to evaluate CF with a large $K$ of hundreds, rather than a small number.}.
From Table~\ref{tab:main}, we have the following key observations: 

1. Our AMF achieves the best results in most cases. The only exception is on Yelp, where IRGAN outperforms AMF by a small margin in NDCG@50 and is on par with AMF in NDCG@100. For the other cases, AMF outperforms other comparing methods statistically significantly with a $p$-value of smaller than $0.01$. This signifies that AMF achieves the state-of-the-art performance for item recommendation. 

2. Specifically, compared to NeuMF --- a recently proposed and very expressive deep learning model, AMF exhibits an average improvement of 2.9\%. This is very remarkable, since AMF uses the shallow MF model that has much fewer parameters, which also implies the potential of improving conventional shallow methods with a better training algorithm. 

3. Moreover, as compared to IRGAN, which also applies adversarial learning on MF but in a different way, AMF betters it by $5.9\%$ on average. This further verifies the effectiveness of our APR method. It is worth mentioning that APR is more efficient and much easier to train than IRGAN, which needs to be carefully tuned to avoid mode collapse, while APR only requires an initialization from BPR. 

4. Among the baselines, NeuMF performs the best, which verifies the advantage of nonlinear neural networks in learning the user-item interaction function. Another neural recommender model CDAE performs weaker, which only shows significant improvements over MF-BPR on the Gowalla dataset. IRGAN manages to outperform MF-BPR in most cases, which can be attributed to its improved training process, since the underlying model is also MF. Lastly, all personalized methods outperform ItemPop by a large margin, which indicates the necessity of personalization in recommendation task. This is not a new finding and has been verified by many previous works~\cite{BPR,iCD,NCF,CDAE,LambdaFM}.

\subsection{Hyper-parameter Studies (RQ3)}
Our APR method introduces two additional hyper-parameters $\epsilon$ and $\lambda$ to control the noise level and the strength of adversarial regularizer, respectively. Here we show how do the two hyper-parameters impact the performance and also shed lights on how to set them. Due to space limitation, we show the results on the Pinterest and Gowalla datasets only, and the results on the Yelp dataset show exactly the same trend.

\begin{figure}[h]
	\centering
	\begin{subfigure}[b]{0.235\textwidth}
		\centering
		\includegraphics[width=\textwidth]{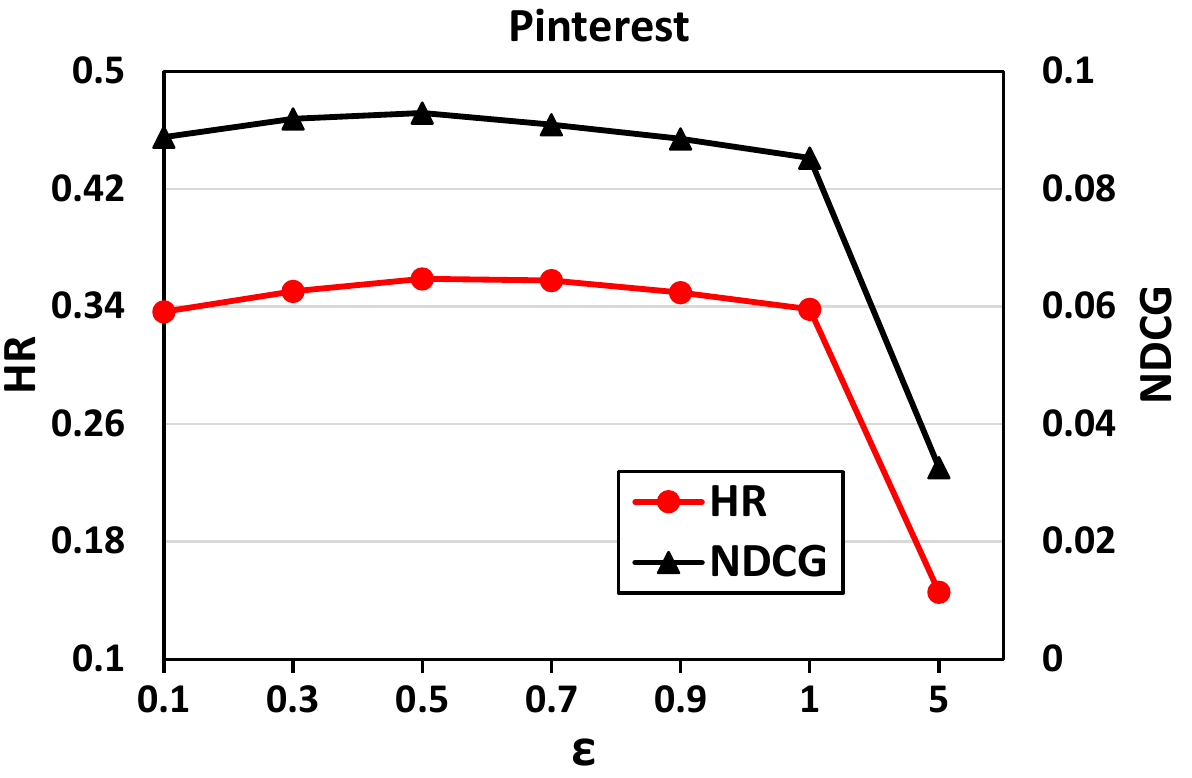}
		\vspace{-18pt}
		\label{fig:pinterest-parameter-eps}
	\end{subfigure} 
	\begin{subfigure}[b]{0.235\textwidth}
		\centering
		\includegraphics[width=\textwidth]{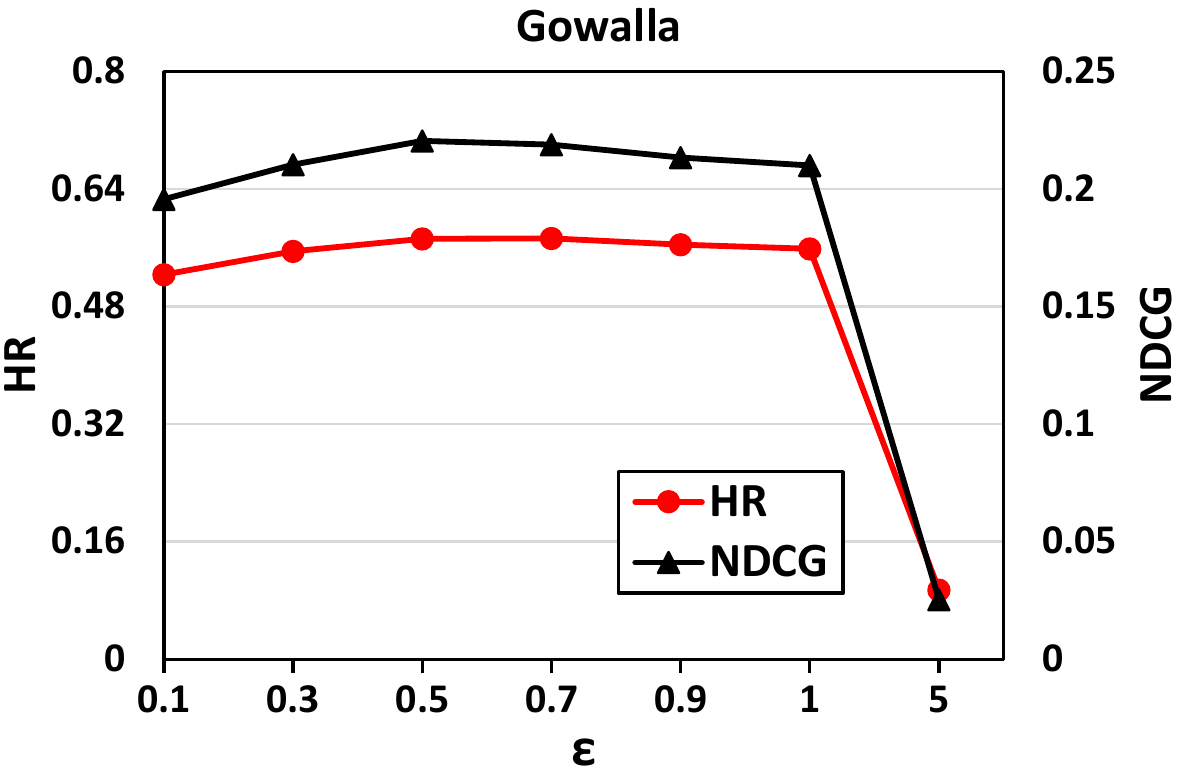}
		\vspace{-18pt}
		\label{fig:gowalla-parameter-eps}
	\end{subfigure}
	\caption{Performance of AMF with respect to different values of $\epsilon$ on Pinterest and Gowalla ($\lambda$ is set to 1).} 
	\vspace{-5pt}
	\label{fig:parameter-eps}
\end{figure}

First, we fix $\lambda$ to the default value of $1$ and vary $\epsilon$. As can be seen from Figure~\ref{fig:parameter-eps}, the optimal value is around $0.5$. When $\epsilon$ is too small (\eg less than 0.1), AMF behaves similarly to MF-BPR and has only minor improvements. This further verifies the positive effect of increasing the robustness of a model to perturbations on its parameters. Moreover, when $\epsilon$ is too large (\eg larger than 1), the performance drops dramatically. This indicates that too large perturbations will destroy the learning process of model parameters. As such, our suggested setting of $\epsilon$ is 0.5 for AMF when it has been pre-trained with BPR. 

\begin{figure}[h]
	\centering
	\begin{subfigure}[b]{0.235\textwidth}
		\centering
		\includegraphics[width=\textwidth]{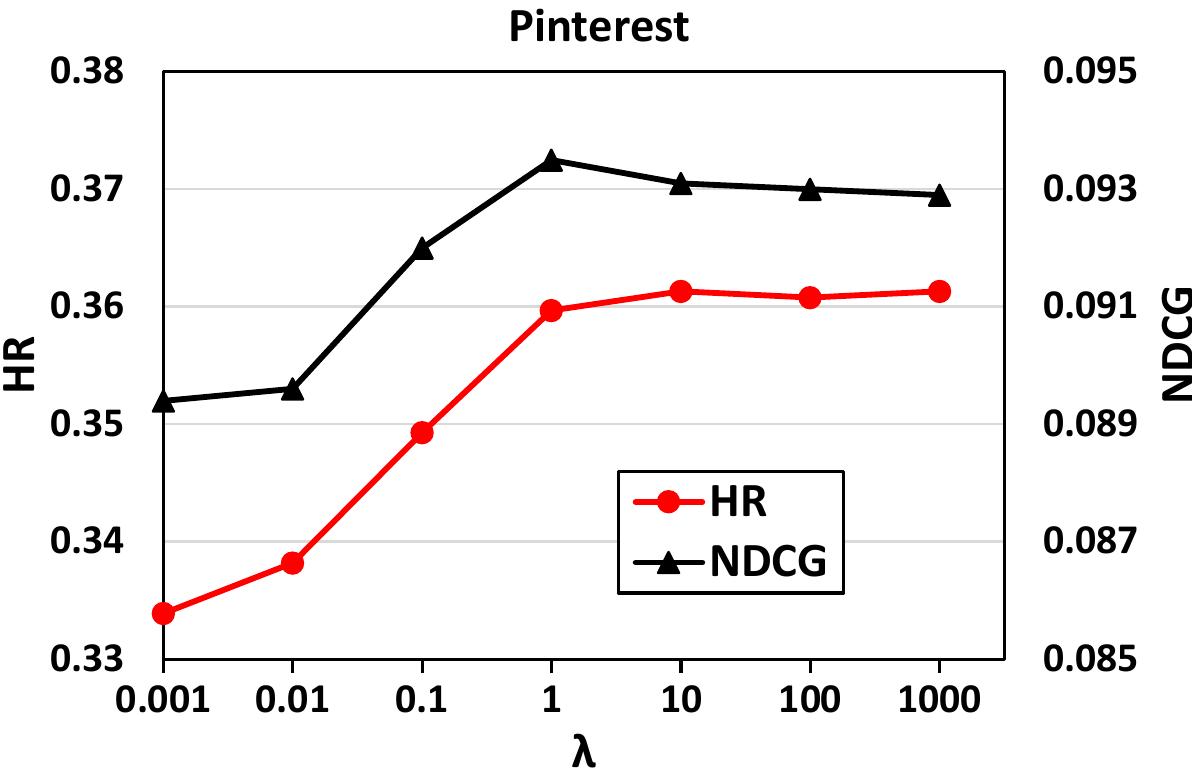}
		\vspace{-18pt}
		\label{fig:pinterest-parameter-lambda}
	\end{subfigure}
	\begin{subfigure}[b]{0.235\textwidth}
		\centering
		\includegraphics[width=\textwidth]{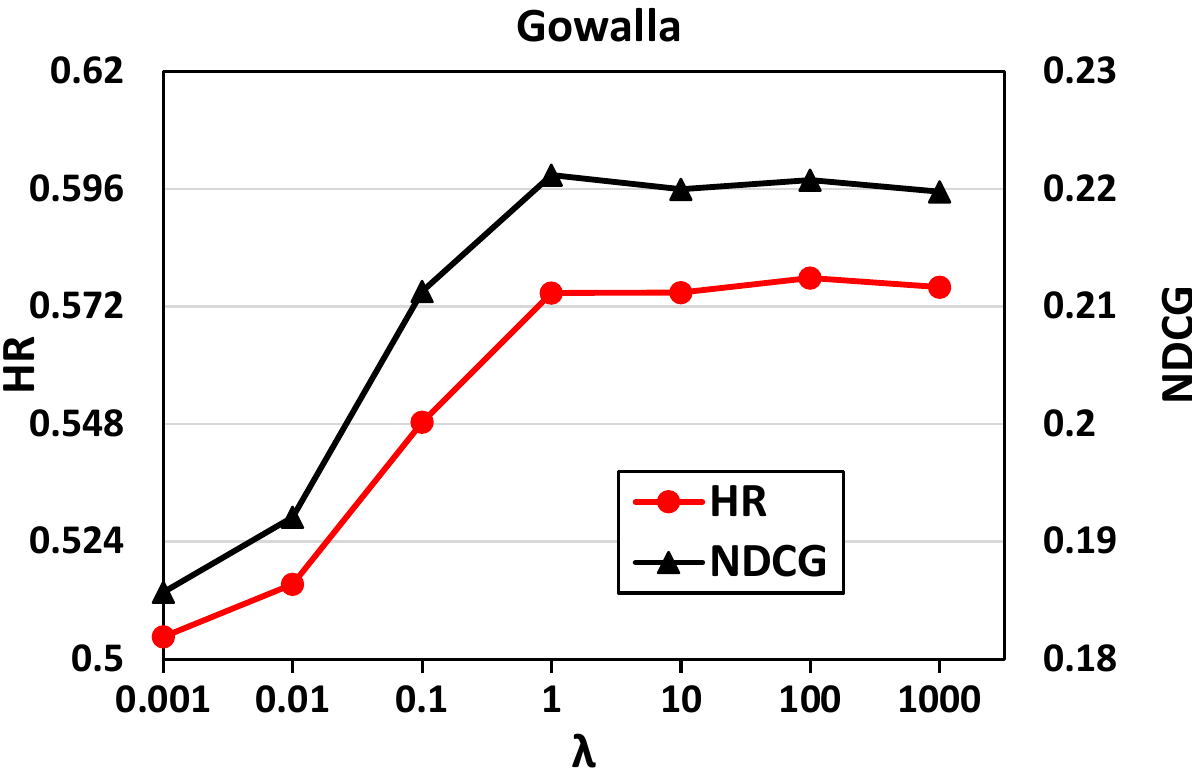}
		\vspace{-18pt}
		\label{fig:gowalla-parameter-lambda}
	\end{subfigure}
	\caption{Performance of AMF with respect to different values of $\lambda$ on Pinterest and Gowalla ($\epsilon$ is set to 0.5).} 	\vspace{-5pt}
	\label{fig:parameter-lambda}
\end{figure}

Second, we fix $\epsilon$ to 0.5 and vary $\lambda$. Figure \ref{fig:parameter-lambda} shows the results. We can see that when $\lambda$ is smaller than 1, increasing $\lambda$ leads to gradual improvements. When $\lambda$ is larger than 1, further increasing it neither improves nor decreases the performance up until a large value of $1,000$. This means that AMF is rather insensitive when $\lambda$ is sufficiently large to reflect the adversarial effect. As such, we suggest to set $\lambda$ to 1 (or a larger value such as 10) for AMF.

\section{Related Work}
\label{sec:related}

\subsection{Item Recommendation}
Due to the abundance of user feedback such ratings and purchases that can directly reflect a user's preference, research on item recommendation have mainly focused on mining the feedback data, known as collaborative filtering (CF). 
Among the various CF methods, matrix factorization (MF)~\cite{fastMF}, a special type of latent factor models, is a basic yet most effective recommender model. 
Popularized by the Netflix Challenge, early works on CF have largely focused on explicit ratings~\cite{SVD++,FM}. 
These works formulated the recommendation task as a regression problem to predict the rating score. 
Later on, some research found that a good CF model in rating prediction may not necessarily perform well in top-K recommendation~\cite{cremonesi2010}, and called on recommendation research to focus more on the ranking task. 

Along another line, research on CF has gradually shifted from explicit ratings to one-class implicit feedback~\cite{BPR,fastMF}. Rendle \etal~\cite{BPR} first argued that item recommendation is a personalized ranking task, and such that, the optimization should be tailored for ranking rather than regression. They then proposed a pairwise learning method BPR, which optimizes a model based on the relative preference of a user over pairs of items. Later on, BPR has been used to optimize a wide range of models~\cite{ACF,JRL,cao2017embedding,Yu:2018:ACR,CDAE,CKBE,LambdaFM}, being a dominant technique in recommendation. 
Recently, Ding \etal~\cite{www2018improvedBPR} improved BPR with a better negative sampler by additionally leveraging view data in E-commerce. 
Our proposed APR directly enhances BPR by adversarial training, having the potential to improve all existing recommender systems based on BPR. 

From the perspective of models, there are many recent efforts developing non-linear neural network models for CF~\cite{CDAE,NCF,cheng2018ijcai,NNCF,ACF,Silkroad,JRL,NAS,Chen:2017,yang2017bridging,LatentCross} to take advantage of deep learning. In particular, He \etal~\cite{NCF} argued the limitation of fixed interaction function (\ie inner product) in MF, and proposed a neural collaborative filtering~(NCF) framework that learns the interaction function from data. They then designed a NCF model named NeuMF, which unifies the strength of MF and MLP in learning the interaction function. Later on, the NCF framework was extended to incorporate the neighborhoods~\cite{NNCF} and attributes~\cite{Silkroad} of users and items, to model contexts for POI recommendation~\cite{yang2017bridging},
to model image/video content features for multi-media recommendation~\cite{ACF}, 
to model aspects in textual reviews~\cite{cheng2018ijcai}, 
to recommend items for a group of users~\cite{cao2018attentive},
and so on. 
In addition to the feed-forward NCF framework, recurrent neural networks have also been developed to
handle the temporal signal in session-aware recommendation~\cite{NAS,LatentCross}. 

\subsection{Adversarial Learning}
This work is inspired by the recent developments of adversarial machine learning techniques~\cite{AML_2014,AML_2015,moosavi2016universal,AdversarialDropout,miyato2016adversarial,ATforRE}. Briefly speaking, it was found that normal supervised training process makes a classier vulnerable to adversarial examples~\cite{AML_2014}, which revealed the potential issue of an unstable model in generalization. To address the issue, researchers then proposed adversarial training methods which augment the training process by dynamically generating adversarial examples~\cite{AML_2015}. Learning over these adversarial examples can be seen as a way to regularize the training process. 
Recently, the idea of adversarial training has been extended to 
learn adaptive dropout for hidden layers in deep neural networks~\cite{AdversarialDropout}. 

Existing work on the emerging field of adversarial learning was largely focused on the domain of image classification. There are very few studies on adversarial learning for ranking --- the core task in IR. The work that is most relevant with ours is IRGAN~\cite{IRGAN}, which also employs adversarial learning, more precisely the GAN framework~\cite{GAN}, to address the matching problem. Our APR methodology is fundamentally different from IRGAN, which aims to unify the strength of generative and discriminative models. 
Specifically, in the pairwise formulation of IRGAN, the generator approximates the relevance distribution to generate document (item) pairs given a query (user), and the discriminator tries to distinguish whether the document pairs are from real data or generated. 
Unfortunately, it is intuitively difficult to understand why IRGAN-pairwise can improve personalized ranking in item recommendation (in fact, both the original paper and their released codes only have IRGAN-pointwise for the recommendation task). 

It is worth noting that in the literature of recommender systems, the concept of \textit{robustness} usually refers to the degree that an algorithm can resist the \textit{profile injection attack}, \ie the attack that tries to manipulate the recommendation by inserting user profiles~\cite{book:robustCF}. This line of research is orthogonal to our work, since we consider improving a recommender model by making it resistant to adversarial perturbations on its parameters. Through this way, we can get a more robust and stable predictive function, and in turn improving its generalization performance. To the best of our knowledge, this has never been explored before in the domain of IR. 

\section{Conclusion and Future Work}
\label{sec:conclusion}

This work contributes a new learning method for optimizing recommender models. 
We show that a model optimized by BPR, a dominant pairwise learning method in recommendation, is vulnerable to adversarial perturbations on its parameters. 
This implies the possible weakness of a model optimized with BPR in generalization. 
Towards the goal of learning more robust models for personalized ranking, we propose to perform adversarial training on BPR, namely, Adversarial Personalized Ranking. 
We develop a generic learning algorithm for APR based on SGD, and employ the algorithm to optimize MF. 
In our evaluation, we perform extensive analysis to demonstrate the highly positive effect of adversarial learning for personalized ranking. 

In future, we plan to extend our APR method to other recommender models. First, we are interested in exploring more generic feature-based models like Neural Factorization Machines~\cite{NFM} and Deep Crossing~\cite{DeepCrossing} that can support a wide range of recommendation scenarios, such as cold-start, context-aware, session-based recommendation and so on. Second, we will employ APR on the recently developed neural CF models such as NeuMF~\cite{NCF} and neighbor-based NCF~\cite{NNCF} to further advance the performance of item recommendation. The challenge here is how to properly employ adversarial training on deep hidden layers, since this work addresses the embedding layer of shallow MF model only. Lastly, it is worth mentioning that our APR represents a generic methodology to improve pairwise learning by using adversarial training. Pairwise learning is not specific to recommendation, and it has been widely applied to many other IR tasks, such as text retrieval, web search, question answering, knowledge graph completion, to name a few. We will work on extending the impact of APR to these fields beyond recommendation. \vspace{+5pt}

\noindent\textbf{Acknowledgments}. 
This work is supported by NExT, by the National Research Foundation Singapore under its AI Singapore Programme, Linksure Network Holding Pte Ltd and the Asia Big Data Association (Award No.: AISG-100E-2018-002), and by the National Natural Science Foundation of China under Grant No.: 61702300.

\bibliographystyle{abbrv}\balance
\bibliography{proc}\balance
\end{document}